\documentclass[utf8]{frontiersSCNS} % for Science, Engineering and Humanities and Social Sciences articles
\makeatletter
\@addtoreset{subfigure}{figure}
\makeatother

\usepackage{url,hyperref,lineno,microtype}
\usepackage[onehalfspacing]{setspace}

\usepackage{listings}
\usepackage{epsfig}
\usepackage{amsmath}
\usepackage{amsfonts}
\usepackage{amsbsy}
\usepackage{mathdots}
\usepackage{mathdots}
\usepackage{lipsum}
\usepackage[utf8]{inputenc}
\usepackage{enumerate}
\usepackage{amsmath}
\usepackage{amssymb}
\usepackage{mathabx}
\usepackage{centernot}
\usepackage[mathscr]{euscript}
\usepackage{url}
\usepackage{hyphenat}
\usepackage{verbatim}
\usepackage{accents}
\usepackage{graphicx}
\usepackage{subcaption}
\usepackage{listings}
\usepackage{graphicx}
\usepackage{grffile}
\usepackage{algorithm,algorithmicx}
\usepackage{algpseudocode}
\algnewcommand\algorithmicinput{\textbf{Input: }}
\algnewcommand\algorithmicoutput{\textbf{Output: }}
\usepackage{afterpage}
\usepackage{epstopdf}
\usepackage{smartdiagram} % https://www.latex4technics.com/?note=1TUS
\usesmartdiagramlibrary{additions}
\usetikzlibrary{shapes.geometric,arrows,quotes}
\usepackage{tikz}

\usepackage[normalem]{ulem}

\everymath{\displaystyle}
\def\keyFont{\fontsize{8}{11}\helveticabold }
\def\firstAuthorLast{Tran {et~al.}} %use et al only if is more than 1 author
\def\Authors{Anh Tran\,$^{1,*}$, Tim Wildey\,$^{1}$, and Hojun Lim\,$^{2}$}
\def\Address{
$^{1}$Scientific Machine Learning, Sandia National Laboratories, Albuquerque, NM 87123, USA, 
$^{2}$Computational Materials and Data Science, Sandia National Laboratories, Albuquerque, NM 87123, USA
}
\def\corrAuthor{Anh Tran}

\def\corrEmail{anhtran@sandia.gov}

\begin{document}
\onecolumn
\firstpage{1}

\title[UQ for CPFEM constitutive models]{Microstructure-sensitive uncertainty quantification for crystal plasticity finite element constitutive models using stochastic collocation methods}

\author[\firstAuthorLast ]{\Authors} %This field will be automatically populated
\address{} %This field will be automatically populated
\correspondance{} %This field will be automatically populated

\extraAuth{}% If there are more than 1 corresponding author, comment this line and uncomment the next one.
%\extraAuth{corresponding Author2 \\ Laboratory X2, Institute X2, Department X2, Organization X2, Street X2, City X2 , State XX2 (only USA, Canada and Australia), Zip Code2, X2 Country X2, email2@uni2.edu}

\maketitle

\begin{abstract}

%%% Leave the Abstract empty if your article does not require one, please see the Summary Table for full details.
\section{}
% For full guidelines regarding your manuscript please refer to \href{http://www.frontiersin.org/about/AuthorGuidelines}{Author Guidelines}.

% As a primary goal, the abstract should render the general significance and conceptual advance of the work clearly accessible to a broad readership. References should not be cited in the abstract. Leave the Abstract empty if your article does not require one, please see \href{http://www.frontiersin.org/about/AuthorGuidelines#SummaryTable}{Summary Table} for details according to article type. 

% \lipsum[2]

Uncertainty quantification (UQ) plays a major role in verification and validation for computational engineering models and simulations, and establishes trust in the predictive capability of computational models. 
In the materials science and engineering context, where the process-structure-property-performance linkage is well known to be the only road mapping from manufacturing to engineering performance, numerous integrated computational materials engineering (ICME) models have been developed across a wide spectrum of length-scales and time-scales to relieve the burden of resource-intensive experiments. 
Within the structure-property linkage, crystal plasticity finite element method (CPFEM) models have been widely used since they are one of a few ICME toolboxes that allows numerical predictions, providing the bridge from microstructure to materials properties and performances. 
Several constitutive models have been proposed in the last few decades to capture the mechanics and plasticity behavior of materials. 
While some UQ studies have been performed, the robustness and uncertainty of these constitutive models have not been rigorously established. 
In this work, we apply a stochastic collocation (SC) method, which is mathematically rigorous and has been widely used in the field of UQ, to quantify the uncertainty of three most commonly used constitutive models in CPFEM, namely phenomenological models (with and without twinning), and dislocation-density-based constitutive models, for three different types of crystal structures, namely face-centered cubic (fcc) copper (Cu), body-centered cubic (bcc) tungsten (W), and hexagonal close packing (hcp) magnesium (Mg). 
Our numerical results not only quantify the uncertainty of these constitutive models in stress-strain curve, but also analyze the global sensitivity of the underlying constitutive parameters with respect to the initial yield behavior, which may be helpful for robust constitutive model calibration works in the future.

\tiny
 \keyFont{ \section{Keywords:} uncertainty quantification, crystal plasticity finite element, constitutive models, stochastic collocation, sparse grid, polynomial chaos expansion} %All article types: you may provide up to 8 keywords; at least 5 are mandatory.
\end{abstract}

\section{Introduction}

% For Original Research Articles \cite{conference}, Clinical Trial Articles \cite{article}, and Technology Reports \cite{patent}, the introduction should be succinct, with no subheadings \cite{book}. For Case Reports the Introduction should include symptoms at presentation \cite{chapter}, physical exams and lab results \cite{dataset}.

% - what is the problem you are solving
% - why is this problem important 
% - what people have done so far to solve this problem
% - what approach do you take to solve the problem (an overview)
% - why is your approach unique / what is your technical contribution / what is new

% \textcolor{red}{emphasize on how PCE becomes a cornerstone UQ method and how wide it has applied to other fields}

Uncertainty quantification (UQ) has been a cornerstone in applied mathematics to verify and validate forward models, with application ranging from subsurface flow, climate change, integrated computational materials engineering (ICME) developments, advanced manufacturing, and many more. 
In the context of materials design, process-structure-property-performance relationship plays a critical role in establishing the linkage between manufacturing and desired properties, to which materials can be tailored for specific applications. 
Across the spectrum of length- and time-scales, from quantum to macro-scale, multiple ICME models have been developed in the quest of accurate prediction of materials properties and performance. 
Notable ICME models for deformation of metals includes, but are not limited to, concurrent or hierarchical couplings of density functional theory, molecular dynamics, kinetic Monte Carlo, dislocation dynamics, crystal plasticity, phase-field, finite element, and any sort of \textcolor{black}{multi-physics and} hybrid approaches.
Despite its success, there are still much room and open questions for UQ research with respect to microstructure-induced mechanical response in metal alloys. 

The concept of ICME models is particularly important because with the predictive capability, engineers can numerically approximate material properties and performance under different operating conditions, without actually performing physical experiments. With the rise of high-performance parallel computing, this step is often hailed as the third paradigm of science~\cite{agrawal2016perspective}, supporting prior empirical and experimental, as well as theoretical research. 
However, as in any forward computational model that relies on deterministic calculations, there is a necessary need to quantify the uncertainty that is associated with the numerical predictions, in order to sufficiently enhance its fidelity and robustness. 
For the case of crystal plasticity finite element method (CPFEM) as an ICME model that bridges microstructure and materials properties and performance, the question of uncertainty is even more relevant because microstructures are well known to be stochastic, high-dimensional in image or volume representations, and may be anisotropic and heterogeneous.

Numerous UQ studies in computational solid mechanics have been conducted over the last decade. 
Given the critical importance of optimization and UQ for a wide variety of problems in materials science, several frameworks have been developed, see e.g., \cite{panchal2013key,mcdowell2007simulation,kalidindi2016vision}, to provide robust predictions under uncertainty.
A comprehensive review of UQ applications in ICME-based simulations can be found in Honarmandi and Arr{\'o}yave \cite{honarmandi2020uncertainty}. 
For example, 
Zhao et al.~\cite{zhao2022quantifying} incorporated measurement and parametric uncertainty to quantify the uncertainty of critical resolved share stress for hcp Ti alloys from nano-indentation. 
Lim et al.~\cite{lim2019investigating} investigated the mesh sensitivity and polycrystalline representative volume element (RVE), where initial textures, hardening models, and boundary conditions are uncertain, and showed that an adequate polycrystalline RVE is obtained by capturing 1000 or more grains. 
Tran and Wildey~\cite{tran2020solving} applied data-consistent inversion method to infer a distribution of microstructure features from a distribution of yield stress, where the push-forward density map via a heteroscedastic Gaussian process approximation is consistent with a target yield stress density. 
Kotha et al.~\cite{kotha2019parametrically1,kotha2019parametrically2,kotha2020uncertainty1,kotha2020uncertainty2} developed uncertainty-quantified parametrically homogenized constitutive models to capture uncertainty in microstructure-dependent stress-strain curve, as well as stochastic yield surface, which has been broadly applied for modeling multi-scale fatigue crack nucleation in Ti alloys~\cite{ozturk2019two,ozturk2019parametrically} and for single-crystal Ni-based superalloys with support vector regression as an underlying machine learning model~\cite{weber2020machine}. 
Sedighiani et al.~\cite{sedighiani2020efficient,sedighiani2022determination} applied genetic algorithm and polynomial approximation to various constitutive models, including phenomenological and dislocation-density-based models. 
Tran et al.~\cite{tran2019quantifying} applied stochastic collocation (SC) method to quantify uncertainty for dendrite morphology and growth via phase-field model. 
Acar et al.~\cite{acar2017stochastic} proposed a linear programming approach to maximize a mean of materials properties under the assumption of Gaussian distribution for both inputs and outputs. 
Fernadez et al.~\cite{fernandez2018estimating} utilized Bayesian inference to quantify the uncertainty in stress-strain curves, where model parameters are treated as random variables. 
Tallman et al.~\cite{tallman2019gaussian,tallman2020uncertainty} applied Gaussian process regression and the Materials Knowledge System framework to predict a set of homogenized materials properties with uncertainty from a distribution function for crystallography orientations and textures.
Inductive design exploration method (IDEM) \cite{ellis2017application,mcdowell2009integrated,choi2008inductive} has been introduced as a materials design methodology to identify feasible and robust design for microstructure features, which has been broadly applied to many practical problems.
From a methodological perspective, numerous applied mathematical UQ techniques have been developed over a few decades. 
While both intrusive and non-intrusive options are available, non-intrusive polynomial chaos expansion (PCE)~\cite{xiu2002wiener,najm2009uncertainty}, which is also known as non-intrusive spectral projection PCE, is one of the most widely used UQ methods to propagate uncertainty in physical models and computational simulations. 
Global sensitivity analysis methods, which is arguably constructed on top of the high-dimensional model representation, have also well studied. 
And apparently, numerous studies connecting PCE and global sensitivity analysis, for example, Sudret~\cite{sudret2008global}, Crestaux et al.~\cite{crestaux2009polynomial}, and Saltelli et al.~\cite{saltelli2010variance}, have been conducted. 
\textcolor{black}{Despite the fact that numerous UQ application works have been carried out in the materials literature and more specifically, in the structure-property relationship, to the knowledge of the authors, none has been applied to quantify the uncertainty related to the underlying constitutive models used in CPFEM.}

In this work, we adopt crystallographic textures from Kocks et al.~\cite{kocks1998texture}(cf. Table I, Chapter 5, pg 185) and ~\cite{wenk2004texture}(Tables 1 and 2), Raabe et al.~\cite{raabe2002theory}, Raabe and Roters~\cite{raabe2004using}, Pham et al.~\cite{pham2017roles}, and Mangal and Holm~\cite{mangal2018dataset}. 
We limit the scope of UQ on the microstructure-mechanical property linkage, where the crystal plasticity finite element is widely regarded as the microstructure-aware multi-scale ICME model, and adopt the DREAM.3D~\cite{groeber2014dream} and DAMASK~\cite{roters2019damask} workflow as previously demonstrated by Diehl et al.~\cite{diehl2017identifying}, as well as two constitutive models described in Sedighiani et al.~\cite{sedighiani2020efficient,sedighiani2022determination}, to perform uncertainty quantification via DAKOTA~\cite{eldred2009recent,dalbey2021dakota}.

\textcolor{black}{The rest of the paper is organized as follows.  
Section~\ref{sec:uqBackground} introduces PCE and SC as the UQ backbone methodology used in this paper. 
Section~\ref{sec:UQworkflowCPFEM} describes the integrated UQ workflow and how they are implemented in practice. 
Section~\ref{sec:PhenoFccCu} shows the first case study of 5d UQ for fcc Cu using phenomenological constitutive model with slipping. 
Section~\ref{sec:PhenoHcpMg} shows the second case study of 16d UQ for hcp Mg using phenomenological constitutive model with slipping and twinning. 
Section~\ref{sec:DisloBccW} shows the last case study of 7d UQ for bcc W using dislocation-density-based constitutive model. 
Section~\ref{sec:discussion} discusses and Section~\ref{sec:conclusion} concludes the paper, respectively. 
}

\section{Uncertainty quantification background}
\label{sec:uqBackground}

In this section, we describe UQ background for CPFEM. 
In Section~\ref{subsec:gPC}, we summarize the theoretical foundation for generalized polynomial chaos expansion as a non-intrusive spectral projection method. 
In Section~\ref{subsec:SC}, we provide the mathematical background for Smolyak sparse grid construction for high-dimensional interpolation and integration, as well as some comparison to full tensor grid highlighting the computational advantage in SC method. 
We refer interested readers to~\cite{babuvska2007stochastic,nobile2008sparse,xiu2009fast} for a more rigorous mathematical characterization of the SC method. 

\subsection{Generalized polynomial chaos expansion}
\label{subsec:gPC}

The generalized Wiener-Askey PCE ~\cite{xiu2002wiener} represents the second-order random process $f(\theta)$ as
\begin{equation}
\label{eq:gPC}
\begin{array}{lll}
f(\theta) & = & c_0 I_0 + \sum_{i_1=1}^{\infty} c_{i_1} I_1(\xi_{i_1}(\theta)) \\
&+& \sum_{i_1=1}^{\infty} \sum_{i_2=1}^{\infty} c_{i_1 i_2} I_2(\xi_{i_1}(\theta), \xi_{i_2}(\theta)) \\
&+& \sum_{i_1=1}^{\infty} \sum_{i_2=1}^{\infty} \sum_{i_3=1}^{\infty} c_{i_1 i_2 i_3} I_3(\xi_{i_1}(\theta), \xi_{i_2}(\theta), \xi_{i_3}(\theta)) + \cdots,
\end{array}
\end{equation}
where $I_n(\xi_{i_1}, \cdots, \xi_{i_n})$ denotes the Wiener-Askey polynomial chaos of order $n$ in terms of the random vector $\boldsymbol{\xi} = (\xi_{i_1}, \xi_{i_2}, \dots, \xi_{i_n})$, and $c$'s are polynomial chaos expansion coefficients to be determined. Without loss of generality, Equation~\ref{eq:gPC} can be rewritten as  
\begin{equation}
\label{eq:shortGPC}
f(\theta) = \sum_{j=0}^{\infty} \widehat{f}_j \boldsymbol{\Phi}_j(\boldsymbol{\xi(\theta)}),
\end{equation}
where there is a one-to-one correspondence between the function $I_n(\xi_{i_1}, \cdots, \xi_{i_n})$ and $\boldsymbol{\Phi}_j(\boldsymbol{\xi})$. 
% Let $\theta$ be the random event in a sample space $\Theta$ with probability measure $P$, and $f(\theta)$ be a second order stochastic process. PCE is a means of representing $f$ parametrically through a set of random variables $\{\xi_i(\theta)\}_{i=1}^d,\, d\in \mathbb{N}$: 
% \begin{equation}
% \label{eq:shortGPC}
% f(\theta) = \sum_{j=0}^{\infty} \widehat{f}_j \boldsymbol{\Phi}_j(\boldsymbol{\xi}(\theta)),
% \end{equation}
% where 
$\boldsymbol{\Phi}_j(\boldsymbol{\xi}(\theta))$ are orthogonal polynomials in terms of $\boldsymbol{\xi} := \{\xi_i(\theta)\}_{i=1}^d$, i.e.,  
\begin{equation}
\langle \boldsymbol{\Phi}_i \boldsymbol{\Phi}_j \rangle = \langle \boldsymbol{\Phi}_j^2 \rangle \delta_{ij},
\end{equation}
where $\delta_{ij}$ is the Kronecker delta and $\langle \cdot, \cdot \rangle$ denotes the weighted-average, which is defined as the inner product in the Hilbert space of the variable $\boldsymbol{\xi}$ with respect to the weighting function $\boldsymbol{W}(\boldsymbol{\xi})$, described in Table~\ref{tab:WienerAskeyScheme}, as
\begin{equation}
\langle \boldsymbol{\Phi}_i \boldsymbol{\Phi}_j \rangle := \int_{\theta \in \Theta}\boldsymbol{\Phi}_i(\boldsymbol{\xi}) \boldsymbol{\Phi}_j(\boldsymbol{\xi})  \boldsymbol{W}(\boldsymbol{\xi}) d \boldsymbol{\xi}.
\end{equation} 
Here, $\widehat{f}_j$ are the coefficients to be determined. 
In practice, the number of terms in \eqref{eq:shortGPC} are truncated after a finite term $P$, where $P+1 = \frac{(p+n)!}{p!n!}$, where $p$ is the order of PCE, and $n$ is the dimensionality of the problem, resulting in an approximation for finite PCE, as
\begin{equation}
\label{eq:finiteShortGPC}
f(\theta) \approx \sum_{j=0}^{P} \widehat{f}_j \boldsymbol{\Phi}_j(\boldsymbol{\xi}).
\end{equation}
The PCE coefficients $\widehat{f}_j$ is determined by non-intrusive spectral projection of \eqref{eq:finiteShortGPC} onto the orthogonal polynomial basis $\{\boldsymbol{\Phi}_j\}$ as
\begin{equation}
\label{eq:pceCoef}
\widehat{f}_j = \frac{\langle f \boldsymbol{\Phi}_j \rangle}{\langle \boldsymbol{\Phi}_j^2 \rangle}.
\end{equation}

Table~\ref{tab:WienerAskeyScheme} describes the relationship between the types of Wiener-Askey polynomial chaos and their corresponding underlying random variables. 
For uniformly distributed variables $\boldsymbol{\xi}$ used in this paper, the Wiener-Askey scheme~\cite{xiu2002wiener} requires Legendre polynomials as the polynomial basis $\{\boldsymbol{\Phi}_j\}$.

\begin{table}[!hbtp]
\centering
\caption{Relationship between the types of Wiener-Askey polynomial chaos and their underlying random variables $\theta$}
\label{tab:WienerAskeyScheme}
\begin{tabular}{lllll} \hline
random variable & probability density function                                                     & polynomial & support range \\ \hline
Gaussian     & $\frac{1}{\sqrt{2\pi}} e^{ -\frac{\theta^2}{2} } $                                  & Hermite      & $(-\infty, \infty)$              \\
uniform      & $\frac{1}{2}$                                                                       & Legendre     & $[-1,1]$               \\
beta         & $\frac{(1-\theta)^\alpha (1+\theta)^\beta}{2^{\alpha+\beta+1} B(\alpha+1, \beta+1)} $    & Jacobi       & $[-1,1]$              \\ 
gamma        & $\frac{\theta^\alpha e^{-\theta}}{\Gamma(\alpha+1)} $                                    & Laguerre     & $[0,\infty)$             \\ \hline
\end{tabular}
\end{table}

\subsection{Stochastic collocation}
\label{subsec:SC}

Sparse grid methods~\cite{novak1996high,novak1997curse,novak1999simple,barthelmann2000high} are a cornerstone in high-dimensional interpolation and integration that have been used in a variety of \textcolor{black}{disciplines}. % applications. 
In concert with the generalized polynomial chaos expansion~\cite{xiu2002wiener} as a non-intrusive spectral projection approach, SC methods~\cite{babuvska2007stochastic,nobile2008sparse,xiu2009fast} are developed to  improve the efficiency of the generalized polynomial chaos expansion on high-dimensional problems using Smolyak sparse grids for integration. 
In the nutshell, the polynomial chaos expansion coefficients are computed based on the sparse grid framework that significantly reduces the effect of the curse-of-dimensionality.  

Following~\cite{nobile2008sparse}, let $i \geq 1$ and $\left\{ \xi^i_1, \dots, \xi^i_{m_i} \right\} \subset [-1, 1]$ be a sequence of abscissas, we begin by introducing the one-dimensional Lagrange interpolation operator as
\begin{equation}
\mathcal{U}^{i}(f) (\xi) = \sum_{j=1}^{m_i} f(\xi^i_j) L^i_j(\xi),
\end{equation}
where $L^i_j(\xi)$ are the Lagrange polynomials of degree $m_i -1$, i.e. $L^i_j(\xi) = \prod^{m_i}_{k=1, k \neq j } \frac{(\xi - x_k^i)}{(\xi^i_j - \xi^i_k)} $. 
The full tensor product formula is perhaps the most straightforward, as
\begin{equation}
\label{eq:full_tensor}
\mathcal{U}^{m_1} \otimes \cdots \otimes \mathcal{U}^{m_{n}} (f) (\mathbf{\xi}) = \sum_{j_1 = 1}^{m_{1}} \cdots \sum_{j_n = 1}^{m_{n}} f(\xi^{i_1}_{j_1},\dots,\xi^{i_n}_{j_n}) \cdot (L^{i_1}_{j_1} \otimes \cdots \otimes L^{i_n}_{j_n}),
\end{equation}
% Here, $ \mathcal{U}^{m_{i}} (f)$ is one-dimensional Lagrange interpolation in the $i$-th dimension with $m = m_i$, so the rule just employs univariate interpolations and then fills up $n$ dimension by dimension. 
which requires $\prod_{i=1}^n m_i$ functions evaluations. 
Although simple, a major drawback of full tensor product is that the total number of points grows very fast in high dimensions. 
Numerous choices of collocation points are possible, such as Gauss-Legendre, Clenshaw-Curtis, Leja, and Gauss-Patterson~\cite{nobile2016convergence}. 
In this work, the weakly-nested Gaussian abscissas (see Section 3.6.2 of ~\cite{dalbey2021dakota} and~\cite{eldred2009comparison}), which are zeros of orthogonal polynomials, are utilized for quadrature.

Let $\mathcal{U}^0 = 0$ and for $i \geq 1$, define 
\begin{equation}
\Delta^{i} = \mathcal{U}^{i} - \mathcal{U}^{i-1},
\end{equation}
the isotropic Smolyak quadrature formula~\cite{smolyak1963quadrature} is given by
\begin{equation}
\mathcal{A}\left(q,n\right) = \sum_{\boldsymbol{i} \leq q} \Delta^{i_1} \otimes \cdots \otimes \Delta^{i_n},
\end{equation}
or equivalently as,  
\begin{equation}
\mathcal{A}\left(q,n\right) = \sum_{q-n+1 \leq |\boldsymbol{i}| \leq q} (-1)^{q-|\boldsymbol{i}|} \cdot \binom{n-1}{q-|\boldsymbol{i}|} \cdot \mathcal{U}^{i_1} \otimes \cdots \otimes \mathcal{U}^{i_n}.
\end{equation}
where $q\ge n$ is an integer denoting the level of the construction~\cite{wasilkowski1995explicit}.
To compute the operator $\mathcal{A}(q,n)$, one needs to evaluate $f$ on the set of points
\begin{equation}
\mathcal{H}(q,n) = \bigcup_{q-n+1\leq |\boldsymbol{i}| \leq q} \left( \mathcal{\boldsymbol{\xi}}^{i_1} \times \cdots \times \mathcal{\boldsymbol{\xi}}^{i_n} \right),
\end{equation}
where $\mathcal{\boldsymbol{\xi}}^i = \{\xi^i_1,\ldots,\xi^i_{m_i}\} \subset [-1,1]$ is the collection of abscissas used by the univariate interpolating operator $\mathcal{U}^i$. This set is a much smaller subset of those required by the full tensor product rule. If the sets are nested, i.e. $\mathcal{\boldsymbol{\xi}}^{i} \subset \mathcal{\boldsymbol{\xi}}^{i+1}$, then $\mathcal{H}(q,n) \subset \mathcal{H}(q+1,n)$.

\begin{table}[!htbp]
\centering
\caption{The number of collocation points used by sparse grid and full tensor grid.}
\label{tab:num_nodes}
\begin{tabular}{c|cc|cc|cc} \hline
Level  & \multicolumn{2}{c}{$n=5$}            & \multicolumn{2}{|c|}{$n=7$}    & \multicolumn{2}{c}{$n=16$}      \\ \cline{2-7} 
$\ell$ & sparse            & full tensor    & sparse            & full tensor    &  sparse            & full tensor    \\ \hline
0            & 1 & 1 & 1 & 1 & 1 & 1       \\
1            & 11 & 243 & 15 & 2187 & 33 & 4.3e+7   \\
2            & 71 & 16807 & 127 & 823543 & 577 & 3.3e+13   \\
3            & 351 & 759375 & 799 & 170859375 & 7105 & 6.5e+18    \\
4            & 1391 & 28629151 & 4047 & 27512614111 & 68865 & 7.2e+23    \\
5            & 4623 & 992436543 & 17263  & 3938980639167 & 556801 & 6.1e+28    \\ \hline
\end{tabular}
\end{table}

To illustrate the benefits in using the Smolyak sparse grid, compared to the full tensor grid, Table~\ref{tab:num_nodes} compares the number of simulations required to achieve the same level of accuracy. The dimensionalities are chosen according to the case studies in this paper. The equivalent number points on full tensor grid point is computed as $(2^{(\ell+1)} - 1)^n$, where $\ell$ is the corresponding level of sparse grid. 

\subsection{Variance-based \textcolor{black}{global} sensitivity analysis}

% \redbf{excerpt Sobol' indices formulation from} ~\cite{saltelli2010variance}
Following~\cite{sudret2008global}, \cite{crestaux2009polynomial}, and~\cite{saltelli2010variance}, we summarize the variance-based \textcolor{black}{global} sensitivity analysis based on Sobol' decomposition as follows. 
In the spirit of generalized polynomial chaos expansion (i.e. Equation~\ref{eq:gPC} after finite truncation), the Sobol' decomposition of $f(\boldsymbol{\xi})$ into the summands of increasing dimensions as
\begin{equation}
\begin{array}{lll}
% f(\xi_1, \dots, \xi_n) &=& f_0 +  \sum_{i=1}^n  \sum_{\alpha \in \mathscr{I}_1} f_\alpha {\boldsymbol{\Phi}}(\xi_i) \\
% && + \sum_{1\leq i_1 < i_2 \leq n} \sum_{\alpha \in \mathscr{I}_{i_1 i_2}} f_\alpha \boldsymbol{\Phi}(\xi_{i_1}, \xi_{i_2}) + \cdots \\
% && + \sum_{1\leq i_1 < \cdots < i_s \leq n} \sum_{\alpha \in \mathscr{I}_{i_1, \dots, i_s}} f_\alpha \boldsymbol{\Phi}(\xi_{i_1}, \dots, \xi_{i_s})  \\
% && + \cdots + \sum_{\alpha \in \mathscr{I}_{1,2,\dots,n}} f_\alpha \boldsymbol{\Phi}(\xi_1,\dots,\xi_n).
f(\xi_1, \dots, \xi_n) &=& \widehat{f}_0 +  \sum_{i=1}^n  \sum_{\alpha \in \mathscr{I}_1} \widehat{f}_\alpha {\boldsymbol{\Phi}}(\xi_i) \\
&& + \sum_{1\leq i_1 < i_2 \leq n} \sum_{\alpha \in \mathscr{I}_{i_1 i_2}} \widehat{f}_\alpha \boldsymbol{\Phi}(\xi_{i_1}, \xi_{i_2}) + \cdots \\
&& + \sum_{1\leq i_1 < \cdots < i_s \leq n} \sum_{\alpha \in \mathscr{I}_{i_1, \dots, i_s}} \widehat{f}_\alpha \boldsymbol{\Phi}(\xi_{i_1}, \dots, \xi_{i_s})  \\
&& + \cdots + \sum_{\alpha \in \mathscr{I}_{1,2,\dots,n}} \widehat{f}_\alpha \boldsymbol{\Phi}(\xi_1,\dots,\xi_n).
\end{array}
\end{equation}

Given a model of the form $y = f(\xi_1, \xi_2, \dots, \xi_n)$, with $y$ as a scalar, a variance-based first order effect for a generic factor $\xi_i$ can be written as $\mathbb{V}_{\xi_i}\left[ \mathbb{E}_{\boldsymbol{\xi}_{\sim i}} \left[ y | \xi_i \right] \right]$, where $\boldsymbol{\xi}_{\sim i}$ is the vector $\boldsymbol{\xi}$ without the $i$-th element, i.e. $\boldsymbol{\xi}_{\sim i} = (\xi_1, \dots, \xi_{i-1}, \xi_{i+1}, \dots, \xi_n)$. 
The main effect sensitivity index (first-order sensitivity coefficient) is written as
\begin{equation}
\label{eq:mainSensitivityIndex}
S_i = \frac{ \mathbb{V}_{\xi_i}\left[ \mathbb{E}_{\boldsymbol{\xi}_{\sim i}} \left[ y | \xi_i \right] \right] }{ \mathbb{V}[y] }.
\end{equation}
It is relatively well-known that 
\begin{equation}
\mathbb{E} \left[ \mathbb{V} \left[ y | \boldsymbol{\xi}_{\sim i} \right] \right] + \mathbb{V} \left[ \mathbb{E} \left[ y | \boldsymbol{\xi}_{\sim i} \right] \right] = \mathbb{V}[y],
\end{equation}
and therefore, the total effect sensitivity index can be obtained as
\begin{equation}
\label{eq:totalSensitivityIndex}
T_i = \frac{ \mathbb{E} \left[ \mathbb{V} \left[ y | \boldsymbol{\xi}_{\sim i} \right] \right] }{ \mathbb{V}[y] } = 1 - \frac{ \mathbb{V} \left[ \mathbb{E} \left[ y | \boldsymbol{\xi}_{\sim i} \right] \right] }{ \mathbb{V}[y] }.
\end{equation} 
\textcolor{black}{In global SA, the importance of parameter $\xi_i$ is measured by comparing its variance of the conditional expectation $\mathbb{V}_{\xi_i}\left[ \mathbb{E}_{\boldsymbol{\xi}_{\sim i}} \left[ y | \xi_i \right] \right]$ against the total variance $\mathbb{V}[y]$. 
}
\textcolor{black}{
$S_i$ measures the effect of $\xi_i$ by evaluating the variance contribution of the basis function $\widehat{f_i}$ that depends strictly on the set of variables in $\xi_i$, while $T_i$ measures the total effect of $\xi_i$ by evaluating the variance contribution of all basis function whose dependencies include $\xi_i$. 
}
For \textcolor{black}{mathematical and implementation details, interested readers are referred to ~\cite{tang2010global}, ~\cite{weirs2012sensitivity},} and ~\cite{crestaux2009polynomial}, where most of the computations are based on Monte Carlo sampling $\boldsymbol{\xi}$. 
In the context of this manuscript, we can understand $\boldsymbol{\xi}$ as the set of parameters for the underlying constitutive model, whether it is phenomenological or dislocation-density-based, and $y$ as the quantity of interest from the CPFEM model. 
\textcolor{black}{It should be emphasized that the method used in this study is the mainstream global sensitivity analysis that is widely used in structural reliability studies.}

\section{Uncertainty quantification workflow for crystal plasticity}
\label{sec:UQworkflowCPFEM}

In this paper, we limit the scope of the UQ studies to cases with a unique RVE.  As discussed in Section~\ref{sec:discussion}, relaxing this restriction will be the subject of future work.  DREAM.3D~\cite{groeber2014dream} is used to generate polycrystalline RVE with a specific crystallographic texture, depending on the material considered. 
DAKOTA~\cite{adams2009dakota} and Python scripts are used to generate inputs for constitutive models, where DAMASK~\cite{roters2019damask} is employed as the CPFEM forward model. 
Results are collected and post-processed in DAKOTA.

% \textcolor{red}{\textbf{insert a schematic picture for the workflow here}}

\begin{figure}[!htbp]
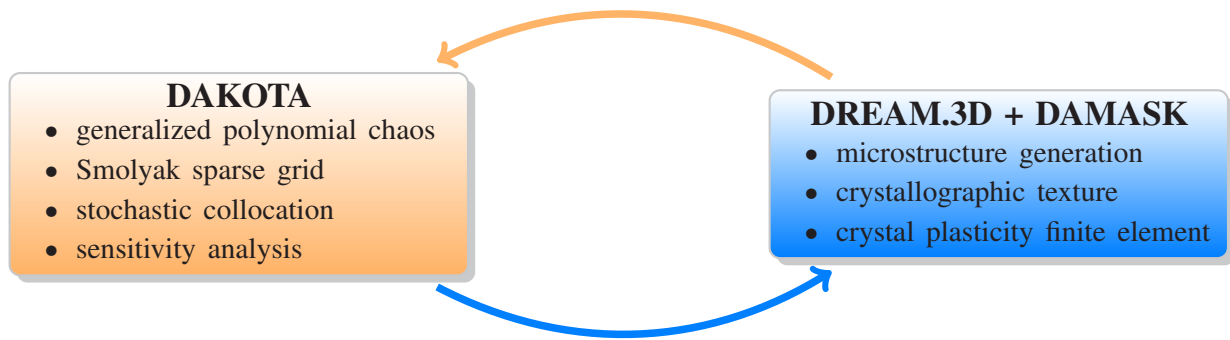

\begin{center}
\smartdiagramset{
set color list={orange!60, blue!50!cyan,magenta!60,green!50!lime!60},
circular distance=5cm,
font=\large,
text width=5.75cm,
module minimum width=2.5cm,
module minimum height=1.5cm,
% arrow line width=5pt,
arrow tip=to,
}
\centering
\smartdiagram[circular diagram]{
% \smartdiagramadd[circular diagram:counterclockwise]{
  \textbf{DAKOTA}  \\
  \begin{itemize}
  \small
    \item generalized polynomial chaos 
    \item Smolyak sparse grid
    \item stochastic collocation
    \item sensitivity analysis
  \normalsize
  \end{itemize}
  ,
  % \textbf{DREAM.3D}  \\
  % \begin{itemize}
  % \small
    % \item stochastic volume element
    % \item crystallographic texture
  % \normalsize
  % \end{itemize}
  \textbf{DREAM.3D + DAMASK}  \\ 
  \begin{itemize}
  \small
    \item microstructure generation
    \item crystallographic texture
    \item crystal plasticity finite element
  \normalsize
  \end{itemize}
}
% {
% above right of module1/return QoI(s), 
% below left of module2/query input,
% % above of module3/QoI(s),
% }

% \smartdiagramconnect{->}{module1/module2}
% \smartdiagramconnect{-to,bend right,shorten >=8pt,shorten <=8pt,"no"{midway,above right,text=black}}{module1/module2} 
% \smartdiagramconnect{->}{module2/module1}
% \smartdiagramconnect{->}{module3/module1}

% \smartdiagramadd[circular diagram:clockwise]{B,C,A}{
% right of module1/Text 1 without arrow, 
% above of module1/Text 2,
% left of module3/Text 3,
% above left of module3/Text 4,
% below left of module2/{Something}}
\end{center}
\caption{Integrating DAKOTA uncertainty quantification workflow to DREAM.3D and DAMASK crystal plasticity finite element simulations. In this framework, DAKOTA queries input parameters to the DREAM.3D + DAMASK automatic workflow, and receives output(s)/quantity(quantities) of interest from DREAM.3D + DAMASK.}
\label{fig:UQworkflowCPFEM}
\end{figure}

Figure~\ref{fig:UQworkflowCPFEM} describes the integrated framework coupling DAKOTA uncertainty quantification code and DREAM.3D + DAMASK workflow. 
Based on the sparse grid construction specifications, such as anisotropic / isotropic, sparse grid level, as well as other sensitivity analysis options, DAKOTA sets up a list of input parameters to be determined and evaluated by the coupled DREAM.3D + DAMASK workflow. 
The sets of simulations are then deployed on high-performance computing systems and evaluated in parallel to accelerate the process. 
Typically, for a fixed input parameter vector, an ensemble of microstructural RVEs are used; however, to reduce the computational cost in this study, we limit the scope of our investigation to one microstructure \textcolor{black}{RVE}. 
It should be noted that, if the initial microstructure is fixed, then DAKOTA would interact directly with DAMASK, and the role of DREAM.3D can be conveniently ignored.

\begin{figure}[!htbp]
\centering
\includegraphics[height=400px]{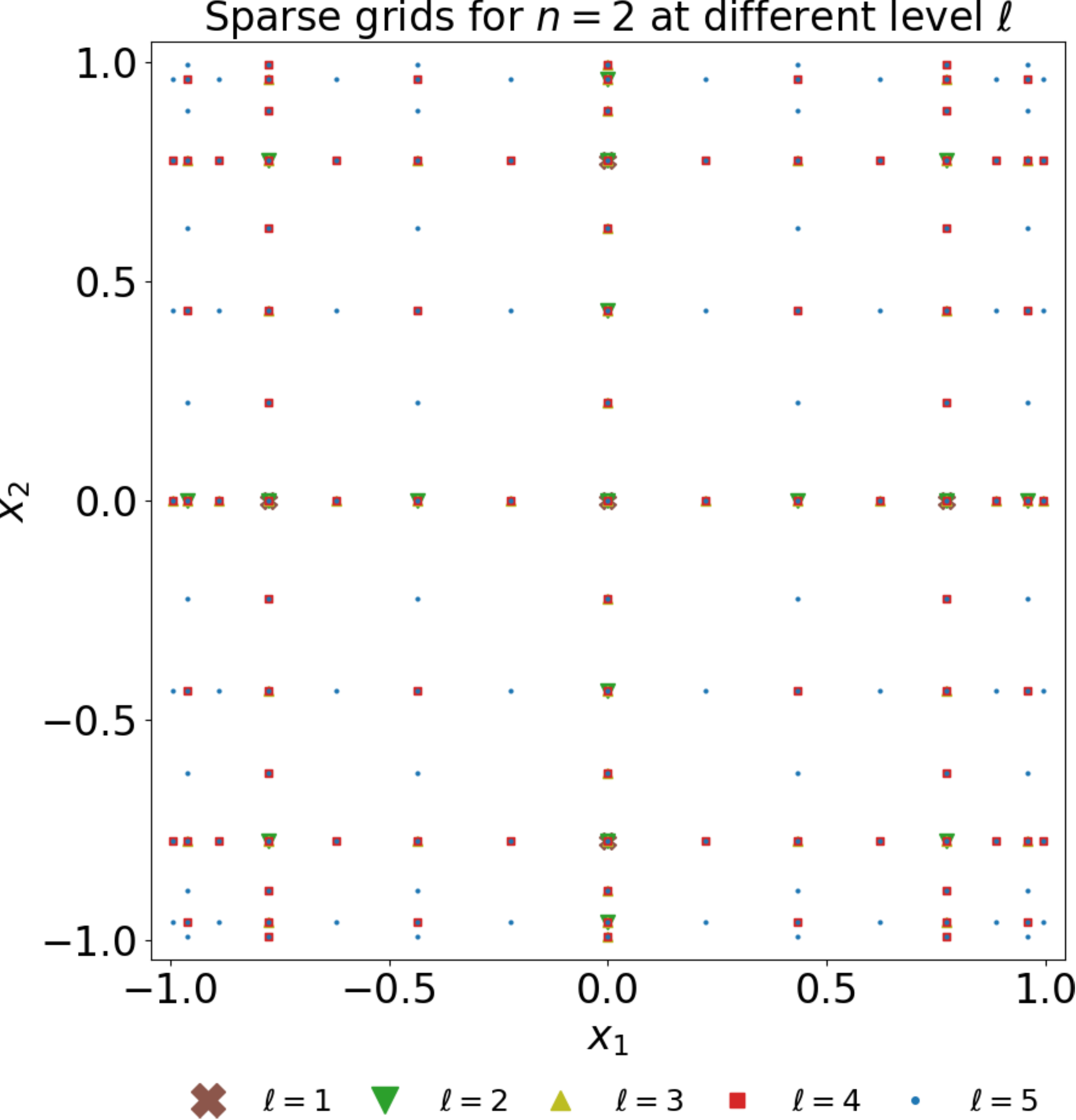}
\caption{Comparison of 2D Smolyak nested sparse grids at various level $\ell$, $1 \leq \ell \leq 5$, with the number of abscissas varies at 5, 17, 49, 97, 161, respectively, using Gaussian abscissas for quadrature.}
\end{figure}

In this section, Section~\ref{subsec:CPFEM} briefly describes the fundamentals of CPFEM model, whereas Section~\ref{subsec:ForwardUQStatement} describes the UQ workflow for CPFEM based on DAKOTA UQ package as the wrapper and DREAM.3D and DAMASK as a forward ICME model.

\subsection{Crystal plasticity model}
\label{subsec:CPFEM}

Consider each point $\mathbf{X}$ in the reference configuration being mapped to the current configuration $\mathbf{x}$ by a linear transformation with the deformation gradient tensor $\mathbf{F}$, where $\mathbf{F} = \frac{\partial \mathbf{x}}{\partial \mathbf{X}}$. 
The Lagrangian strain tensor is defined as
\begin{equation}
\mathbf{E} = 
% \frac{1}{2} (\mathbf{C} - \mathbf{I}) = 
\frac{1}{2} \left( \mathbf{F}^\top \mathbf{F} - \mathbf{I} \right). 
\end{equation}
The total deformation gradient $\mathbf{F}$ can be multiplicatively decomposed into an elastic and plastic parts, 
\begin{equation}
\mathbf{F} = \mathbf{F}_\text{e} \mathbf{F}_\text{p}.
\end{equation}
The velocity gradient, which measures the deformation rate, is defined as
\begin{equation}
\mathbf{L} = \dot{\mathbf{F}} \mathbf{F}^{-1} = \dot{\mathbf{F}}_\text{e} \mathbf{F}_\text{e}^{-1} + \mathbf{F}_\text{e} \left(\dot{\mathbf{F}}_\text{p} \mathbf{F}_\text{p}^{-1}\right)\mathbf{F}_\text{e}^{-1} = \mathbf{L}_\text{e} + \mathbf{F}_\text{e} \mathbf{L}_\text{p} \mathbf{F}_\text{e}^{-1},
\end{equation}
where $\mathbf{L}_\text{p}$ is the plastic velocity gradient evaluated in the intermediate configuration. 
The second Piola-Kirchoff stress measure $\mathbf{S}$ is defined as
\begin{equation}
\mathbf{S} = \mathbb{C} : \mathbf{E}_\text{e} = \frac{\mathbb{C}}{2} \left( \mathbf{F}_\text{e}^\top \mathbf{F}_\text{e} - \mathbf{I} \right),
\end{equation}
where $\mathbf{E}_\text{e}$ is the elastic Green-Lagrange strain tensor, and $\mathbb{C}$ is the fourth-order stiffness tensor. 
% \chhighlight{[DEFINE $\mathbf{C}$ AND $\mathbb{C}]$}
The plasticity velocity gradient $\mathbf{L}_\text{p}$, driven by the second Piola-Kirchoff $\mathbf{S}$, controls the evolution of the plastic deformation gradient as
\begin{equation}
\dot{\mathbf{F}}_\text{p} = \mathbf{L}_\text{p} \mathbf{F}_\text{p}.
\end{equation}

Constitutive 
% \chreplaced{equations representing the flow stress}{models}, 
equations representing the flow stress, 
such as the phenomenological 
% \chadded{slip-based hardening} 
slip-based hardening 
model and the dislocation density-based 
% \chadded{hardening}
hardening 
model, differ on $\mathbf{L}_\text{p}$ are calculated based on a specific microstructure and a set of internal state variables. 
\textcolor{black}{The grain size $d$ in the RVE is a random variable characterized by a log-normal distribution, i.e.,
\begin{equation}
p_D(d; \mu_D, \sigma_D) = \frac{1}{d \sigma_D \sqrt{2\pi}} \exp\left({ - \frac{(\ln d - \mu_D)^2}{2\sigma_D^2} }\right),
\label{eq:GrainLogNormalDistribution}
\end{equation}
where $\mu_D$ and $\sigma_D$ are materials-dependent parameters. 
}

\subsection{Forward uncertainty quantification problem}
\label{subsec:ForwardUQStatement}

In this work, we consider a forward UQ problem for a fixed set (or ensemble) of microstructures using the SC method, where the set of internal state variables for constitutive model parameters are considered stochastic with some inherent uncertainty. 
While the non-intrusive PCE approach allows an arbitrary probability density of parameters, in this work, uniform distributions are imposed on the constitutive model parameters due to lack of prior knowledge. 
With the choice of uniform distributions on bounded intervals, according to the Wiener-Askey scheme shown in Table~\ref{tab:WienerAskeyScheme}, Legendre polynomials are used to approximate the quantities of interest. 

In the initial yield regime, we focus on the the estimated yield strain $\varepsilon_\text{Y}$ and yield stress $\sigma_\text{Y}$.
From the homogenized stress-strain curve $\varepsilon_{vM}-\sigma_{vM}$ obtained from a CPFEM simulation, 
the monotonic cubic interpolation via PCHIP method~\cite{fritsch1984method} is utilized to interpolate $\varepsilon_\text{vM}-\sigma_\text{vM}$ curve. 
From the approximated stress-strain curve, an estimation of modulus of elasticity is obtained by simple linear regression. 
An offset at $\varepsilon = 0.002$ with the estimated modulus of elasticity is drawn, where the coordinates of the intersection are $(\varepsilon_\text{Y}, \sigma_\text{Y})$. 
Statistics of these two quantities of interest are obtained and returned to DAKOTA package. 

To set up the UQ study, a pre-processing compilation of constitutive model parameters are obtained from DAKOTA using a numerical toy model. 
The sets of constitutive model parameters are then appropriately parsed into DAMASK using Python scripts, along with the output of DREAM.3D for setting up the geometric file. 
With the correct setup of DAMASK simulations, the set of DAMSASK simulations are then performed in a massively parallel manner on a high-performance computing cluster. 
The post-processing results are then collected from DAMASK, and parsed back to DAKOTA package using a Python interface. 
DAKOTA then performs the UQ and sensitivity analysis, concluding the UQ workflow for CPFEM based on DREAM.3D and DAMASK. 
% Quantities of interest, in the scope of this paper, include the estimated yield strain $\varepsilon_\text{Y}$, as well as the estimated yield stress $\sigma_\text{Y}$. 
% In order to obtain these quantities, 

\section{Phenomenological constitutive model with slipping in fcc Cu}
\label{sec:PhenoFccCu}

\subsection{Constitutive law}

We adopt the summary and tabulated parameters from~\cite{sedighiani2020efficient,sedighiani2022determination} (Tables 1 and 2). In the phenomenological constitutive model, the shear on each slip system $\alpha$ is modeled as
% \chadded{[IS THIS CORRECT TABLE?]}. 
\begin{equation}
\dot{\gamma}^{\alpha} = \dot{\gamma}_0 \left| \frac{\tau^\alpha}{\tau_0^\alpha}  \right|^n \text{sgn}(\tau^\alpha),
\end{equation}
where $\tau_0$ is the slip resistance, $\dot{\gamma}_0$ is the reference shear rate, and $n$ determines the strain rate sensitivity of slip. 
% \chadded{[MAY BE $n$ IS MISSING IN (18)?]}
The influence of other slip system $\alpha'$ on the hardening behavior of the slip system $\alpha$ is modeled as
\begin{equation}
\dot{\tau}_0^{\alpha} = \sum_{\alpha'=1}^{N_s} h_{\alpha \alpha'} \left|\dot{\gamma}^{\alpha'}\right|,
\end{equation}
where $h_{\alpha \alpha'}$ is the hardening matrix, which captures the micromechanical interaction among different slip systems
\begin{equation}
h_{\alpha \alpha'} = q_{\alpha \alpha'} \left[ h_0 \left( 1 - \frac{\tau_0^\alpha}{\tau_\infty}\right)^a \right],
\end{equation}
$h_0$, $a$, and $\tau_\infty$ are slip hardening parameters for all 12 slip systems in fcc materials. 
$q_{\alpha \alpha'}$ is a measure for latent hardening with value of 1.0 for coplanar slip system $\alpha$ and $\alpha'$ and 1.4 otherwise.

\begin{table}[!htbp]
\tiny
\caption{Parameters for Cu used in this case study (cf. Table 4~\cite{roters2019damask} and Tables 1 and 2~\cite{sedighiani2020efficient,sedighiani2022determination}).}
\label{tab:CuConstitutiveParameters}
\begin{tabular*}{\textwidth}{c @{\extracolsep{\fill}} cccccc} \hline
variable       & description                & units     &  reference value  &  nature       &  distribution       \\ \hline
$C_{11}$       & elastic constant           &   GPa     &  168.3        &  deterministic  &  --             \\   
$C_{12}$       & elastic constant           &   GPa     &  122.1        &  deterministic  &  --             \\
$C_{44}$       & elastic constant           &   GPa     &   75.7        &  deterministic  &  --             \\
$\dot{\gamma}_0$   & reference shear rate   &  s$^{-1}$   &  0.003        &  deterministic  &  --             \\ \hline
$\tau_0$       & slip resistance            &   MPa     &  1.5         &  stochastic     &  $\mathcal{U}[0.5,3.5]$   \\
$\tau_\infty$    & saturation stress        &   MPa     &  112.5        &  stochastic     &  $\mathcal{U}[80,140]$   \\
$h_0$        & slip hardening parameter     &   MPa     &  240        &  stochastic     &  $\mathcal{U}[200,280]$   \\
$n$        & strain rate sensitivity parameter  &   --    &  83.3         &  stochastic     &  $\mathcal{U}[50,120]$  \\
$a$        & slip hardening parameter       &   --    &  2.0         &  stochastic     &  $\mathcal{U}[1,4]$     \\ \hline
\end{tabular*}
\normalsize
\end{table}

\subsection{Design of numerical experiments}

\begin{subfigure}
  \begin{minipage}{0.25\textwidth}
      \centering  
      \includegraphics[width=\linewidth]{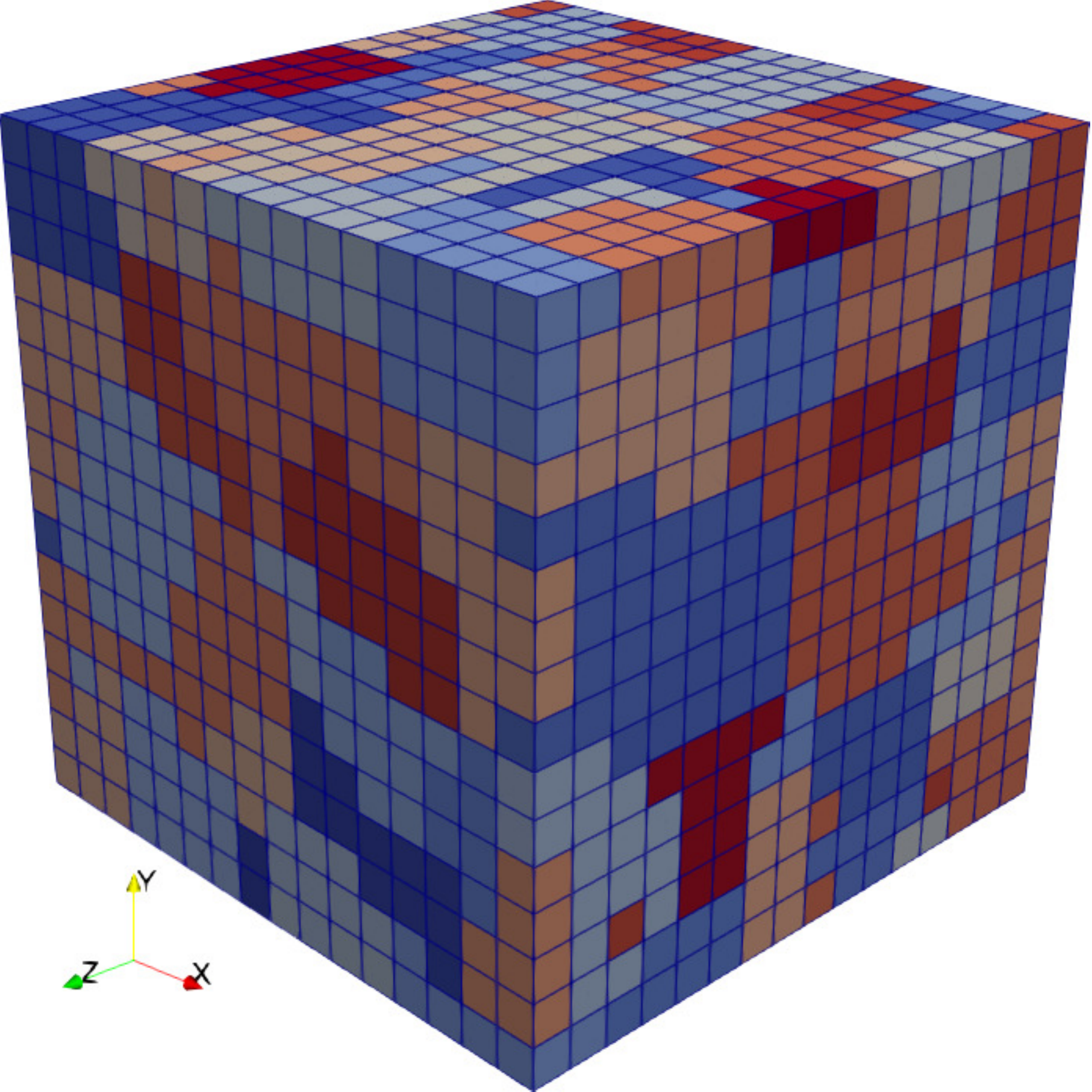}
      \caption{Representative volume element for Cu.}
      \label{fig:copperPhenomenologicalRVE}
    \end{minipage}\hfill%
    % \end{subfigure}
    \begin{minipage}{0.70\textwidth}
    % \begin{subfigure}
      \centering
      \includegraphics[width=\linewidth]{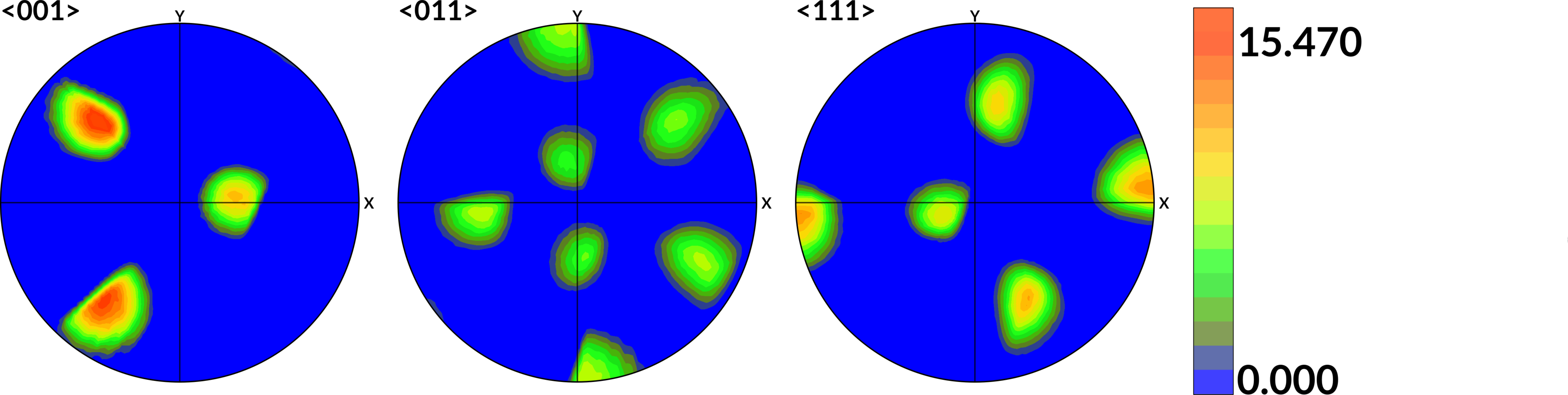}
      \caption{Copper-type of texture component with Euler angles $(\phi_1, \theta, \phi_2) = (90^\circ,35^\circ,45^\circ)$.}
      \label{fig:copperODF}
    % \end{subfigure}
  \end{minipage}% 
% \caption{Microstructure volume element and texture crystallography used in the phenomenological case study.}
\label{fig:phenomenologicalCuMs}
\end{subfigure}

\begin{figure}
\centering
\includegraphics[width=\textwidth]{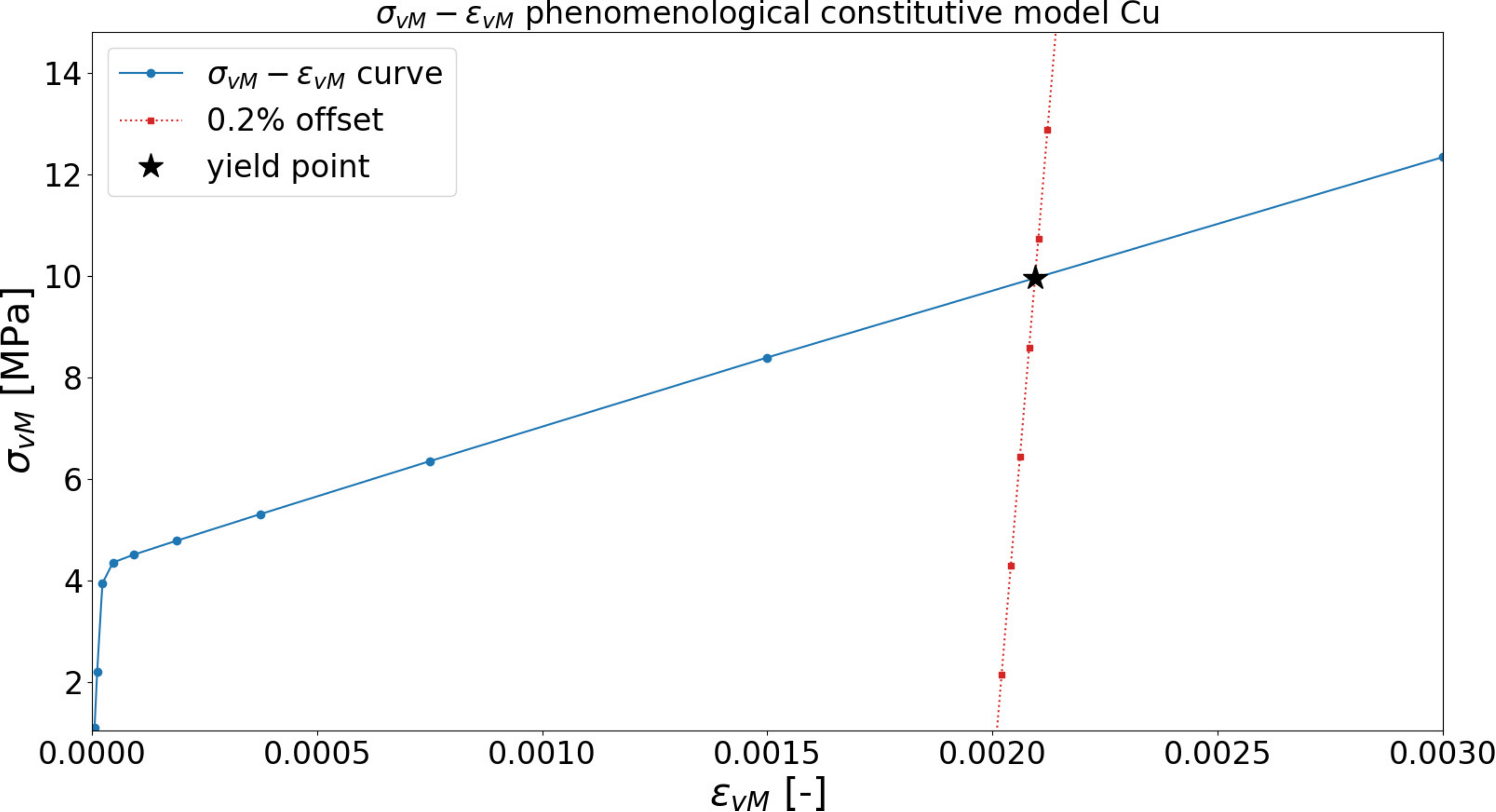}
\caption{Stress-strain equivalent curve for Cu with default parameters and determination of yield point. Modulus of elasticity is estimated as 188.5919 GPa, while $(\varepsilon_\text{Y}, \sigma_\text{Y})$ are estimated as (0.002052, 9.8527 MPa)}
\label{fig:cropped_stress-strain-Cu}
\end{figure}

We conducted our analysis with one RVE shown in Figure~\ref{fig:copperPhenomenologicalRVE}, where the crystallographic texture is shown in Figure~\ref{fig:copperODF}. Average grain size of 60.9467$\mu$m is used, where an RVE of size 256$\mu$m$^3$ is generated. \textcolor{black}{A finite element mesh} of 16$^3$ is created to approximate the microstructure RVE. 
In DREAM.3D, the texture crystallography of Copper-type with Euler angles $(\phi_1, \theta, \phi_2) = (90^\circ,35^\circ,45^\circ)$ is used, the grain size parameters are set as $\mu_D = 4.09$, $\sigma_D = 0.2$, which results in a RVE with 182 grains, shown in Figure~\ref{fig:copperPhenomenologicalRVE}. 
For DAKOTA, we set the sparse grid level $\ell=3$, dimensionality $n=5$, which results in 351 inputs. 
For each set of input parameters, a CPFEM simulation is performed, followed by the post-process. The results are analyzed in the following section. 
A uniaxial loading condition is applied in the [100] direction with $\dot{\varepsilon} = 10^{-3} \text{s}^{-1}$.

To compare with the default parameter, a single CPFEM simulation is performed with constitutive parameters described in Table~\ref{tab:CuConstitutiveParameters}. 
The stress-strain equivalent curve is shown in Figure~\ref{fig:cropped_stress-strain-Cu}. 
% https://www.mit.edu/~6.777/matprops/copper.htm % referred by DAMASK
As a reference to experimental data\footnote{\href{https://www.matweb.com/search/DataSheet.aspx?MatGUID=9aebe83845c04c1db5126fada6f76f7e}{https://www.matweb.com/search/DataSheet.aspx?MatGUID=9aebe83845c04c1db5126fada6f76f7e}}, 
% http://www.matweb.com/search/datasheet.aspx?matguid=9aebe83845c04c1db5126fada6f76f7e&ckck=1
modulus of elasticity for polycrystalline Cu is reported at 110 GPa, whereas its yield strength is reported as 33.3 MPa. 
Compared to the experiment, the computed modulus and yield strength $\sigma_\text{Y}$ are on the same scale, but off roughly by a factor of 1.5$\times$, possibly due to different processing conditions resulting in slightly different alloys.

% \begin{figure}
% \begin{subfigure}{0.20\textwidth}
% \includegraphics[width=\textwidth]{figsSC-CPFEM/cropped_copper_rve-eps-converted-to.pdf}
% \caption{blah}
% \end{subfigure}%
% \hfill
% \begin{subfigure}{0.75\textwidth}
% \includegraphics[width=\linewidth]{figsSC-CPFEM/cropped_copperODF-eps-converted-to.pdf}
% \caption{blah}
% \end{subfigure}

\subsection{Numerical results}

\begin{figure}[!hbtp]
\centering
\includegraphics[width=\textwidth, keepaspectratio]{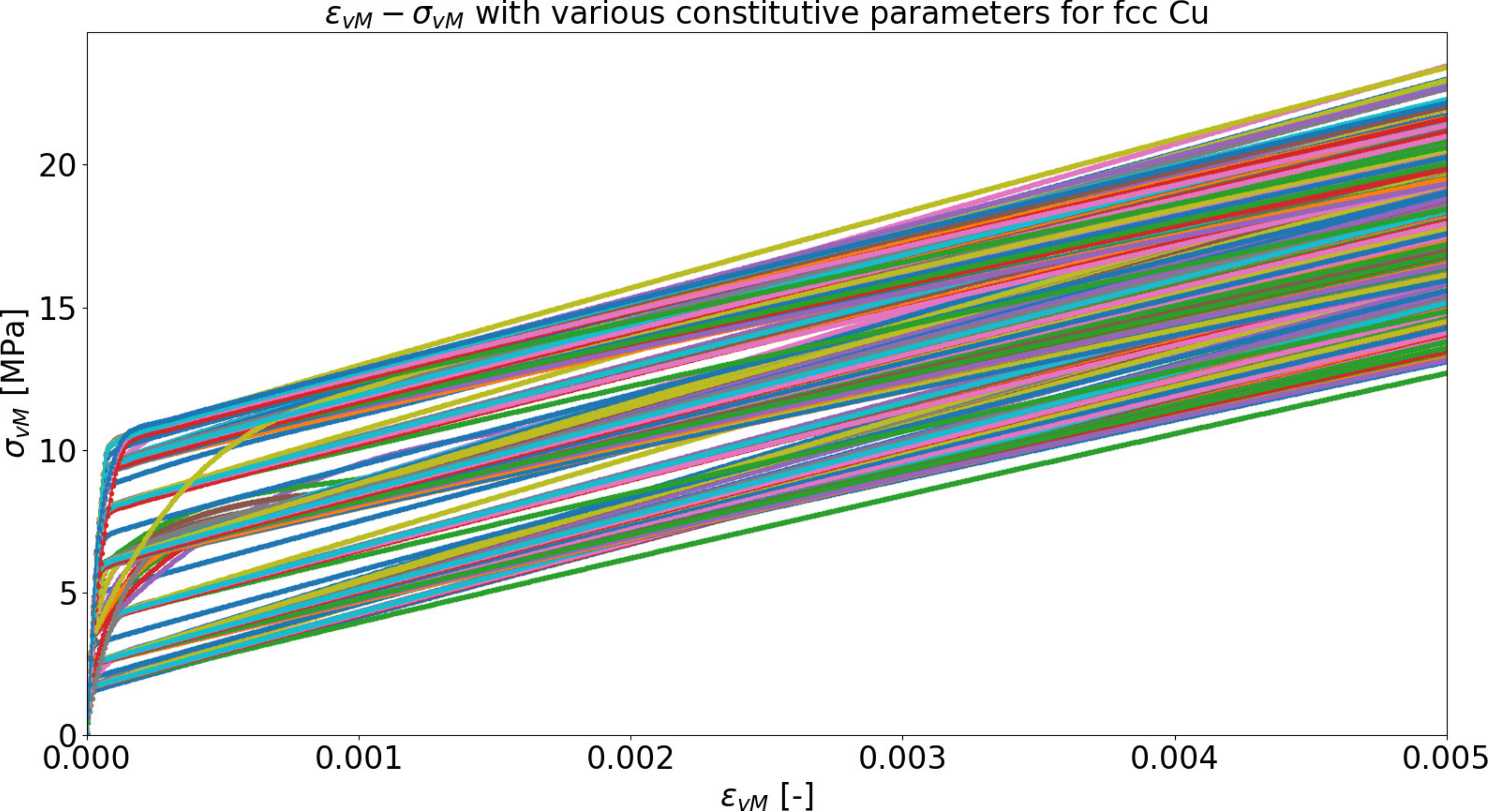}
\caption{Equivalent $\varepsilon_{\text{vM}}-\sigma_{\text{vM}}$ plots for fcc Cu.}
\label{fig:cropped_compiled-stress-strain-Cu-eps-converted-to.pdf}
\end{figure}

Figure~\ref{fig:cropped_compiled-stress-strain-Cu-eps-converted-to.pdf} shows a compilation of stress-strain curve for 351 simulations, where each corresponds to a unique set of constitutive parameters for fcc Cu. As shown in this figure, the constitutive model effects not only the initial yield behavior, but also the modulus of elasticity and the hardening behavior. 

% \begin{figure}[!htbp]
% \centering
% \includegraphics[width=\textwidth, keepaspectratio]{figsSC-CPFEM/cropped_jointPlot-Cu-eps-converted-to.pdf}
% \caption{Joint probability density plot $\varepsilon_{\text{Y}}-\sigma_{\text{Y}}$ for fcc Cu.}
% \label{fig:cropped_jointPlot-Cu-eps-converted-to.pdf}
% \end{figure}

Figures~\ref{fig:cropped_pdf-strainYield-Cu-eps-converted-to.pdf} and \ref{fig:cropped_pdf-stressYield-Cu-eps-converted-to.pdf} show the probability density function for $\varepsilon_{\text{Y}}$ and $\sigma_{\text{Y}}$ , respectively. 
The mode for $\varepsilon_{\text{Y}}$ is approximately 0.00207, whereas the mode for $\sigma_{\text{Y}}$ is approximately 11.65 MPa. 
It should be noted that with the current sparse grid level $\ell = 3$, the approximation for $\varepsilon_\text{Y}$ may be imprecise. 
One of the possible reasons is that the elastic regime of copper is very small (as shown in Figure~\ref{fig:cropped_compiled-stress-strain-Cu-eps-converted-to.pdf}), and therefore, a more accurate approximation may be required to accurately capture the yield strain.

\begin{subfigure}
\setcounter{figure}{6}
\setcounter{subfigure}{0}
  \begin{minipage}{0.475\textwidth}
      \centering  
      \includegraphics[width=\textwidth, keepaspectratio]{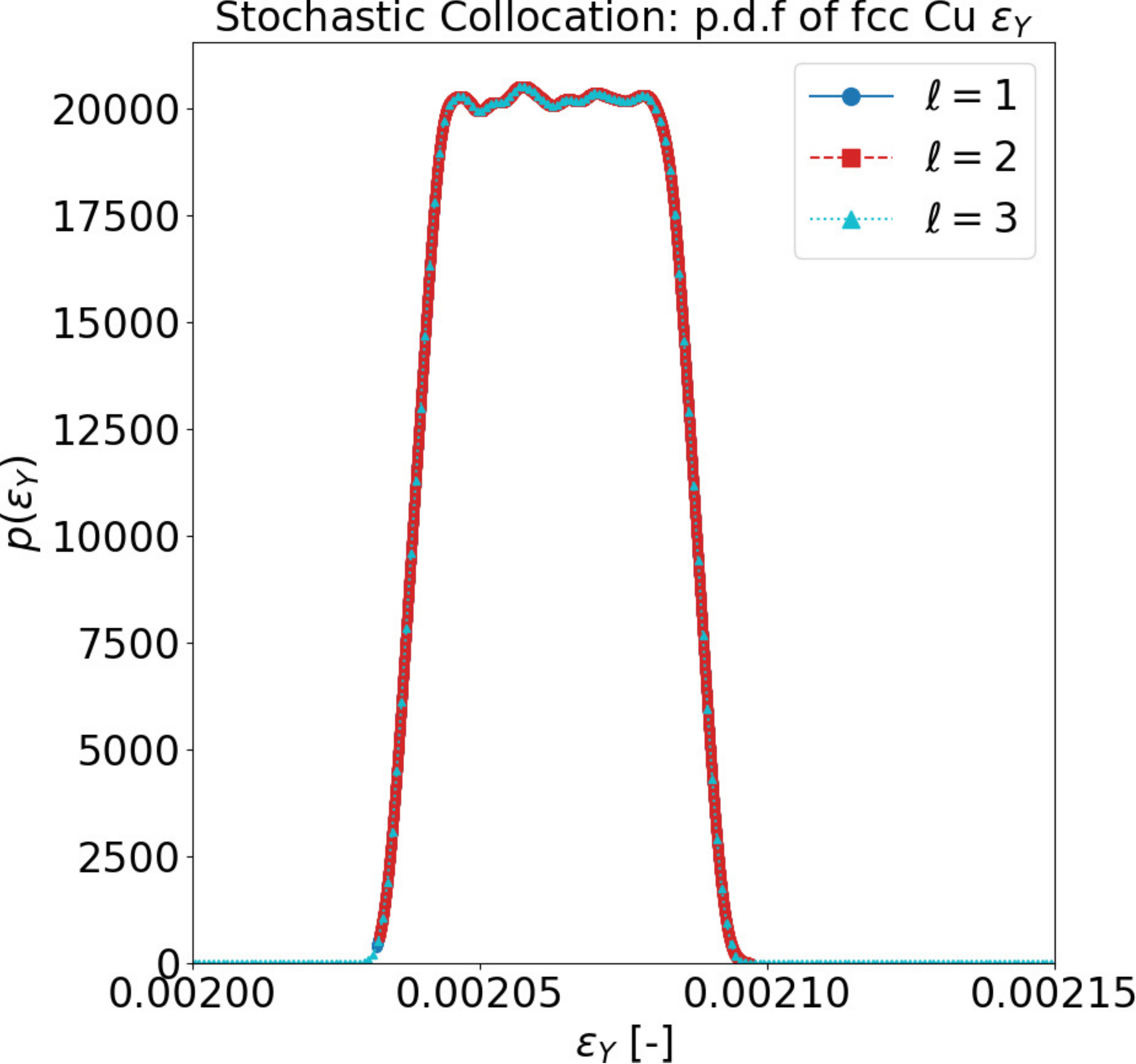}
      \caption{SC probability density function of $\varepsilon_{\text{Y}}$ for fcc Cu.}
      \label{fig:cropped_pdf-strainYield-Cu-eps-converted-to.pdf}
    \end{minipage}\hfill%
    \begin{minipage}{0.475\textwidth}
      \centering
      \includegraphics[width=\textwidth, keepaspectratio]{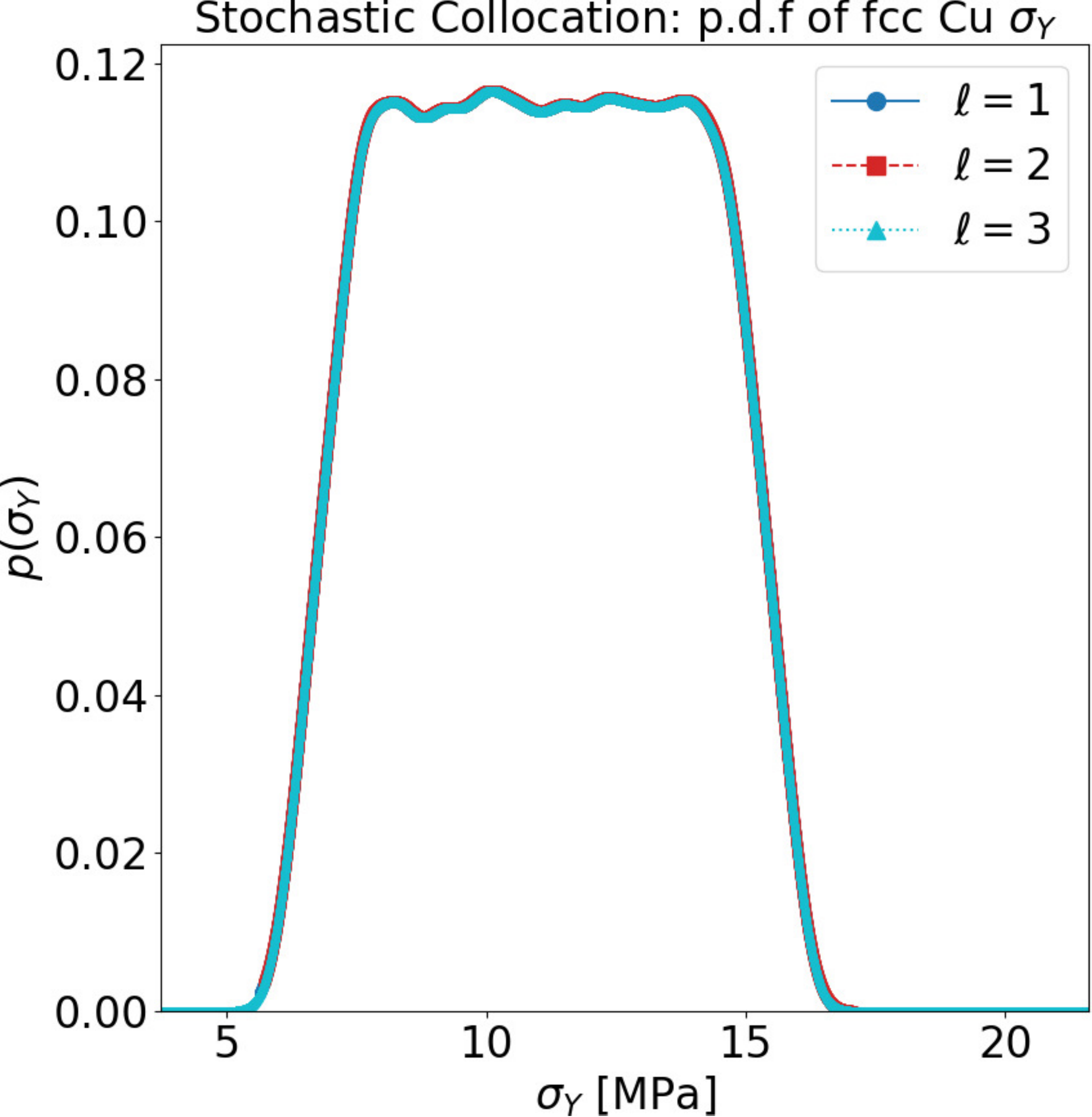}
      \caption{SC probability density function of $\sigma_{\text{Y}}$ for fcc Cu.}
      \label{fig:cropped_pdf-stressYield-Cu-eps-converted-to.pdf}
  \end{minipage}% 
\label{fig:pdf-Cu}
\end{subfigure}

\begin{figure}[!htbp]
\centering
\includegraphics[width=\textwidth, keepaspectratio]{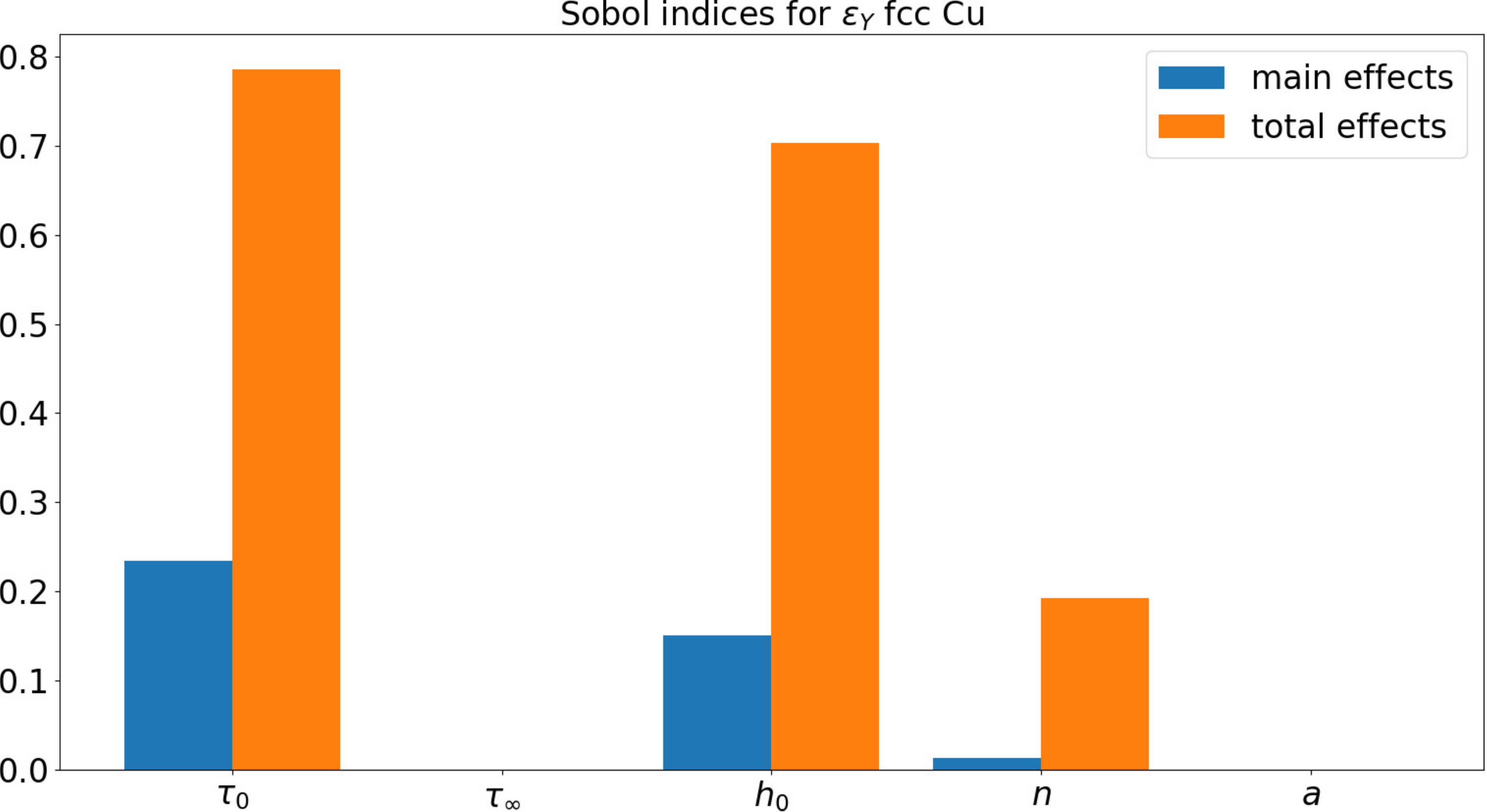}
\caption{Sobol' indices for $\varepsilon_{\text{Y}}$ for fcc Cu.}
\label{fig:cropped_sobol-strainYield-Cu-eps-converted-to.pdf}
\end{figure}

\begin{figure}[!htbp]
\centering
\includegraphics[width=\textwidth, keepaspectratio]{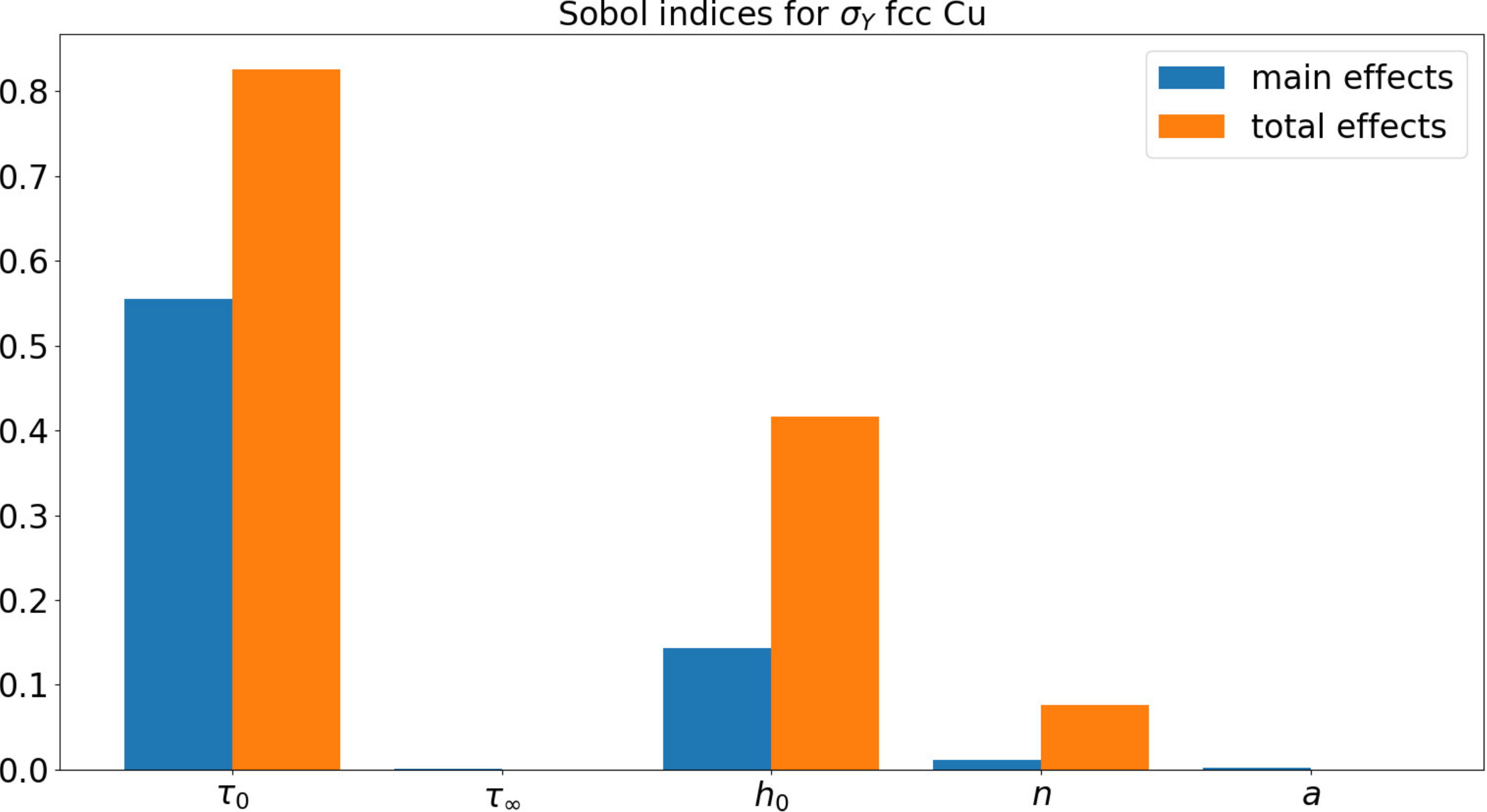}
\caption{Sobol' indices for $\sigma_{\text{Y}}$ for fcc Cu.}
\label{fig:cropped_sobol-stressYield-Cu-eps-converted-to.pdf}
\end{figure}

Figure~\ref{fig:cropped_sobol-strainYield-Cu-eps-converted-to.pdf} and Figure~\ref{fig:cropped_sobol-stressYield-Cu-eps-converted-to.pdf}, respectively, show the Sobol' indices for $\varepsilon_\text{Y}$ and $\sigma_\text{Y}$. 
Ranking from the most influential parameters to the least influential parameters for $\varepsilon_\text{Y}$ from the Sobol indices for main effects, $T_{\tau_0} = 0.7858$, $T_{h_0} = 0.7035$, $T_n = 0.1922$, $T_a = -0.01038$, and $T_{\tau_\infty} = -0.001983$. 
Ranking from the most influential parameters to the least influential parameters for $\sigma_\text{Y}$ from the Sobol indices for main effects, $T_{\tau_0} = 0.8258$, $T_{h_0} = 0.4194$, $T_n = 0.07659$, $T_a = -0.01091$, and $T_{\tau_\infty} = -0.0009990$. 
The order of influential parameters for fcc Cu, regarding the initial yield behavior, is $\tau_0 > h_0 > n > a > \tau_\infty$. 
Since the main scope of this paper is about the initial yield behavior, it is not surprising that Figure~\ref{fig:cropped_sobol-stressYield-Cu-eps-converted-to.pdf} agrees with Figure~\ref{fig:cropped_sobol-strainYield-Cu-eps-converted-to.pdf} in terms of Sobol' indices.

\section{Phenomenological constitutive models with dislocation slip and deformation twinning in hcp Mg}
\label{sec:PhenoHcpMg}

\subsection{Constitutive law}

Firstly introduced by Hutchinson~\cite{hutchinson1976bounds} and extended for twinning by Kalidindi~\cite{kalidindi1998incorporation}, the resistance on $\alpha=1,\dots,N_{\text{s}}$ slip systems evolve from $\xi_0$ to a system-dependent saturation value and depend on shear on slip and twin systems according to
\begin{equation}
\dot{\xi}^{\alpha} = h_0^{\text{s-s}} \left( 1 + c_1 \left(f^{\text{tot}}_{\text{tw}}\right)^{c_2} \right) (1 + h^{\alpha}_{\text{int}}) \left[ \sum_{\alpha' = 1}^{N_{\text{s}}} \left| \dot{\gamma}^{\alpha'} \right| \left| 1 - \frac{\xi^{\alpha'}}{\xi^{\alpha'}_{\infty}} \right|^a \text{sgn}\left( 1 - \frac{\xi^{\alpha'}}{\xi^{\alpha'}_{\infty}} \right) h^{\alpha \alpha'} \right] + \sum_{\beta'=1}^{N_{\text{tw}}} \dot{\gamma}^{\beta'} h^{\alpha \beta'} ,
% \dot{\xi}^{\alpha} = h_0^{\text{s-s}} (1 + h^{\alpha}_{\text{int}}) \left[ \sum_{\alpha' = 1}^{N_{\text{s}}} \left| \dot{\gamma}^{\alpha'} \right| \left| 1 - \frac{\xi^{\alpha'}}{\xi^{\alpha'}_{\infty}} \right|^a \text{sgn}\left( 1 - \frac{\xi^{\alpha'}}{\xi^{\alpha'}_{\infty}} \right) h^{\alpha \alpha'} \right],
\end{equation}
where 
% $f^{\text{tot}}_{\text{tw}}$ is the total twin volume fraction, 
$h$ denotes the components of the slip-slip and slip-twin interaction matrices,
$h_0^{\text{s-s}}$, $h_{\text{int}}$, $c_1$, $c_2$ are model-specific fitting parameters and $\xi_{\infty}$ represents the saturated resistance evolution.

The resistances on the $\beta = 1, \dots, N_{\text{tw}}$ twin systems evolve in a similar way,
\begin{equation}
\dot{\xi}^{\beta} = h_0^{\text{tw-s}} \left( \sum_{\alpha=1}^{N_{\text{s}}} |\gamma_{\alpha}| \right)^{c_3} \left( \sum_{\alpha'=1}^{N_{\text{s}}} \left| \dot{\gamma}^{\alpha'} \right| h^{\beta \alpha'} \right)
+ h_0^{\text{tw-tw}} \left(f^{\text{tot}}_{\text{tw}} \right)^{c_4} \left( \sum_{\beta'=1}^{N_{\text{tw}}} \dot{\gamma^{\beta'}} h^{\beta \beta'} \right),
\end{equation}
where $h_0^{\text{tw-s}}$, $h_0^{\text{tw-tw}}$, $c_3$, and $c_4$ are model-specific fitting parameters. 
Shear on each slip system evolves at a rate of
\begin{equation}
\dot{\gamma}^{\alpha} = (1 - f^{\text{tot}}_{\text{tw}}) \dot{\gamma_0}^{\alpha} \left| \frac{\tau^{\alpha}}{\xi^{\alpha}} \right|^n \text{sgn}(\tau^{\alpha}).
\end{equation}
where slip due to mechanical twinning accounting for the unidirectional character of twin formation is computed slightly differently,
\begin{equation}
\dot{\gamma} = (1 - f^{\text{tot}}_{\text{tw}}) \dot{\gamma_0} \left| \frac{\tau}{\xi} \right|^n \mathcal{H}(\tau),
\end{equation}
where $\mathcal{H}$ is the Heaviside step function. 
The total twin volume is calculated as
\begin{equation}
f^{\text{tot}}_{\text{tw}} = \max\left(1.0, \sum_{\beta=1}^{N_{\text{tw}}} \frac{\gamma^{\beta}}{\gamma^{\beta}_{\text{char}}} \right),
\end{equation}
where $\gamma_{\text{char}}$ is the characteristic shear due to mechanical twinning and depends on the twin system. 
Interested readers are referred to Section 6.2.2 from Roters et al~\cite{roters2019damask}.

\begin{table}[!htbp]
\tiny
\caption{Parameters for Mg used in this case study (cf. Tables 7 and 8~\cite{sedighiani2020efficient,sedighiani2022determination}, ~\cite{wang2014situ,tromans2011elastic,agnew2006validating}).}
\label{tab:MgConstitutiveParameters}
\begin{tabular*}{\textwidth}{c @{\extracolsep{\fill}} cccccc} \hline
variable                    & description                       & units     &  reference value  &  nature       &  distribution         \\ \hline
$c/a$                       & lattice parameter ratio                 &   --    &  1.635        &  deterministic  &  --             \\   
$C_{11}$                    & elastic constant                    &   GPa     &   59.3        &  deterministic  &  --             \\   
$C_{12}$                    & elastic constant                    &   GPa     &   61.5        &  deterministic  &  --             \\
$C_{44}$                    & elastic constant                    &   GPa     &   16.4        &  deterministic  &  --             \\
$C_{44}$                    & elastic constant                    &   GPa     &   25.7        &  deterministic  &  --             \\
$C_{44}$                    & elastic constant                    &   GPa     &   21.4        &  deterministic  &  --             \\ 
$\dot{\gamma}_0$                & twinning reference shear rate             &  s$^{-1}$   &  0.001        &  deterministic  &  --             \\
$\dot{\gamma}_0$                & slip reference shear rate               &  s$^{-1}$   &  0.001        &  deterministic  &  --             \\ \hline
$\tau_{0,\text{basal}}$             & basal slip resistance                 &   MPa     &  10         &  stochastic     &  $\mathcal{U}[5,30]$    \\
$\tau_{0,\text{pris}}$              & prismatic slip resistance               &   MPa     &  55         &  stochastic     &  $\mathcal{U}[30,60]$     \\
$\tau_{0,\text{pyr} \langle a \rangle}$     & pyramidal $\langle a \rangle$ slip resistance     &   MPa     &  60         &  stochastic     &  $\mathcal{U}[50,90]$     \\
$\tau_{0,\text{pyr} \langle c+a \rangle}$     & pyramidal $\langle c+a \rangle$ slip resistance     &   MPa     &  60         &  stochastic     &  $\mathcal{U}[50,110]$    \\
$\tau_{0,\text{T}1}$              & tensile twin resistance                 &   MPa     &  45         &  stochastic     &  $\mathcal{U}[35,70]$     \\
$\tau_{0,\text{C}2}$              & compressive twin resistance               &   MPa     &  80         &  stochastic     &  $\mathcal{U}[60,120]$    \\
$\tau_{\infty,\text{basal}}$          & basal saturation stress                 &   MPa     &  45         &  stochastic     &  $\mathcal{U}[30,60]$     \\
$\tau_{\infty,\text{pris}}$           & prismatic saturation stress               &   MPa     &  135        &  stochastic     &  $\mathcal{U}[100,160]$   \\
$\tau_{\infty,\text{pyr} \langle a \rangle}$  & pyramidal $\langle a \rangle $ saturation stress    &   MPa     &  150        &  stochastic     &  $\mathcal{U}[120,180]$   \\
$\tau_{\infty,\text{pyr} \langle c+a \rangle}$  & pyramidal $\langle c+a \rangle$ saturation stress   &   MPa     &  150        &  stochastic     &  $\mathcal{U}[120,180]$   \\
$h_{0}^{\text{tw}-\text{tw}}$          & twin-twin hardening parameter             &   MPa     &  50          &  stochastic     &  $\mathcal{U}[30,80]$     \\
$h_{0}^{\text{s}-\text{s}}$          & slip-slip hardening parameter             &   MPa     &  500         &  stochastic     &  $\mathcal{U}[100,200]$   \\
$h_{0}^{\text{tw}-\text{s}}$           & twin-slip hardening parameter             &   MPa     &  150         &  stochastic     &  $\mathcal{U}[400,680]$   \\
$n_\text{s}$                  & slip strain rate sensitivity parameter        &   --    &  10         &  stochastic     &  $\mathcal{U}[15,35]$     \\
$n_\text{tw}$                   & twinning strain rate sensitivity parameter      &   --    &  5          &  stochastic     &  $\mathcal{U}[3,8]$     \\
$a$                       & slip hardening parameter                &   --    &  2.5        &  stochastic     &  $\mathcal{U}[2,4]$     \\ \hline
\end{tabular*}
\normalsize
\end{table}

\subsection{Design of numerical experiments}

Similar to the previous section we restrict the scope to one RVE shown in Figure~\ref{fig:magnesiumPhenomenologicalRVE}, where the crystallographic texture is shown in Figure~\ref{fig:magenesiumODF}. Average grain size of 204.037$\mu$m is used, where an RVE of size 2048$\mu$m$^3$ is generated. A mesh of 64$^3$ is created to approximate the microstructure RVE. 
In DREAM.3D, the texture crystallography with Euler angles $(\phi_1, \theta, \phi_2) = (90^\circ,0^\circ,0^\circ)$ is used, the grain size parameters are set as $\mu_D = 5.2983$, $\sigma_D = 0.2$, which results in a RVE with 1706 grains, shown in Figure~\ref{fig:magnesiumPhenomenologicalRVE}. 
For DAKOTA, we set the sparse grid level $\ell=2$, dimensionality $n=16$, which results in 577 inputs. 
For each set of input parameters, a CPFEM simulation is performed, followed by the post-processing steps. The results are analyzed in the following section. 
A uniaxial loading condition is applied in the [100] direction with $\dot{\varepsilon} = 10^{-3} \text{s}^{-1}$.

\begin{subfigure}
\setcounter{figure}{9}
\setcounter{subfigure}{0}
  \begin{minipage}{0.25\textwidth}
      \centering  
      \includegraphics[width=\linewidth]{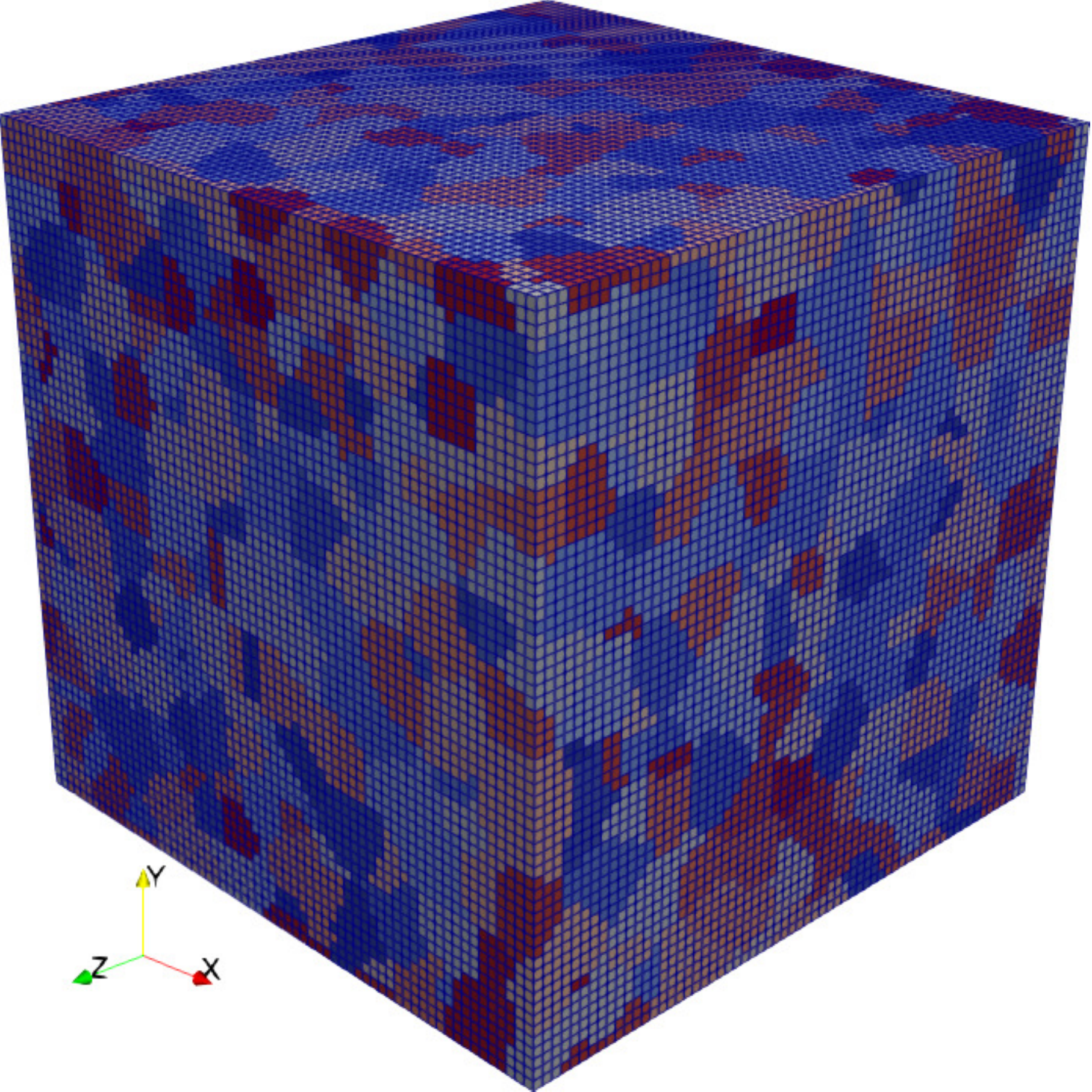}
      \caption{Representative volume element for Mg.}
      \label{fig:magnesiumPhenomenologicalRVE}
    \end{minipage}\hfill%
    \begin{minipage}{0.70\textwidth}
      \centering
      \includegraphics[width=\linewidth]{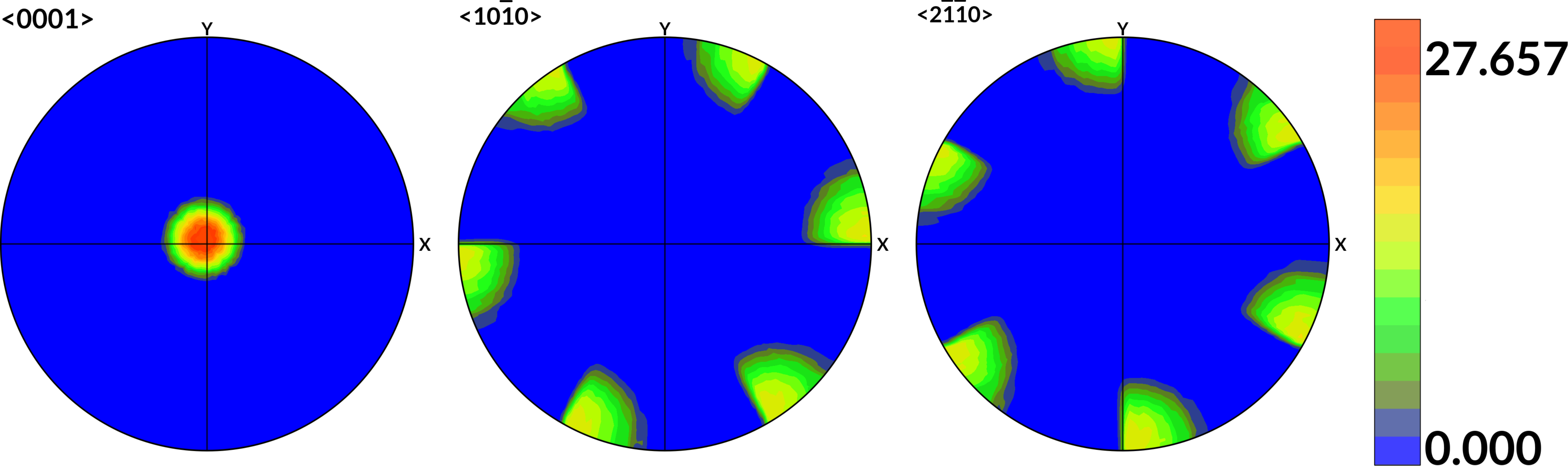}
      \caption{Magnesium texture component with Euler angles $(\phi_1, \theta, \phi_2) = (90^\circ,0^\circ,0^\circ)$ ~\cite{mangal2018dataset}.
      % Sahoo et al.~\cite{sahoo2020strain}.
      }
      \label{fig:magenesiumODF}
  \end{minipage}% 
% \caption{Microstructure volume element and texture crystallography used in the phenomenological case study.}
\label{fig:phenomenologicalMgMs}
\end{subfigure}%

\begin{figure}
\centering
\includegraphics[width=\textwidth]{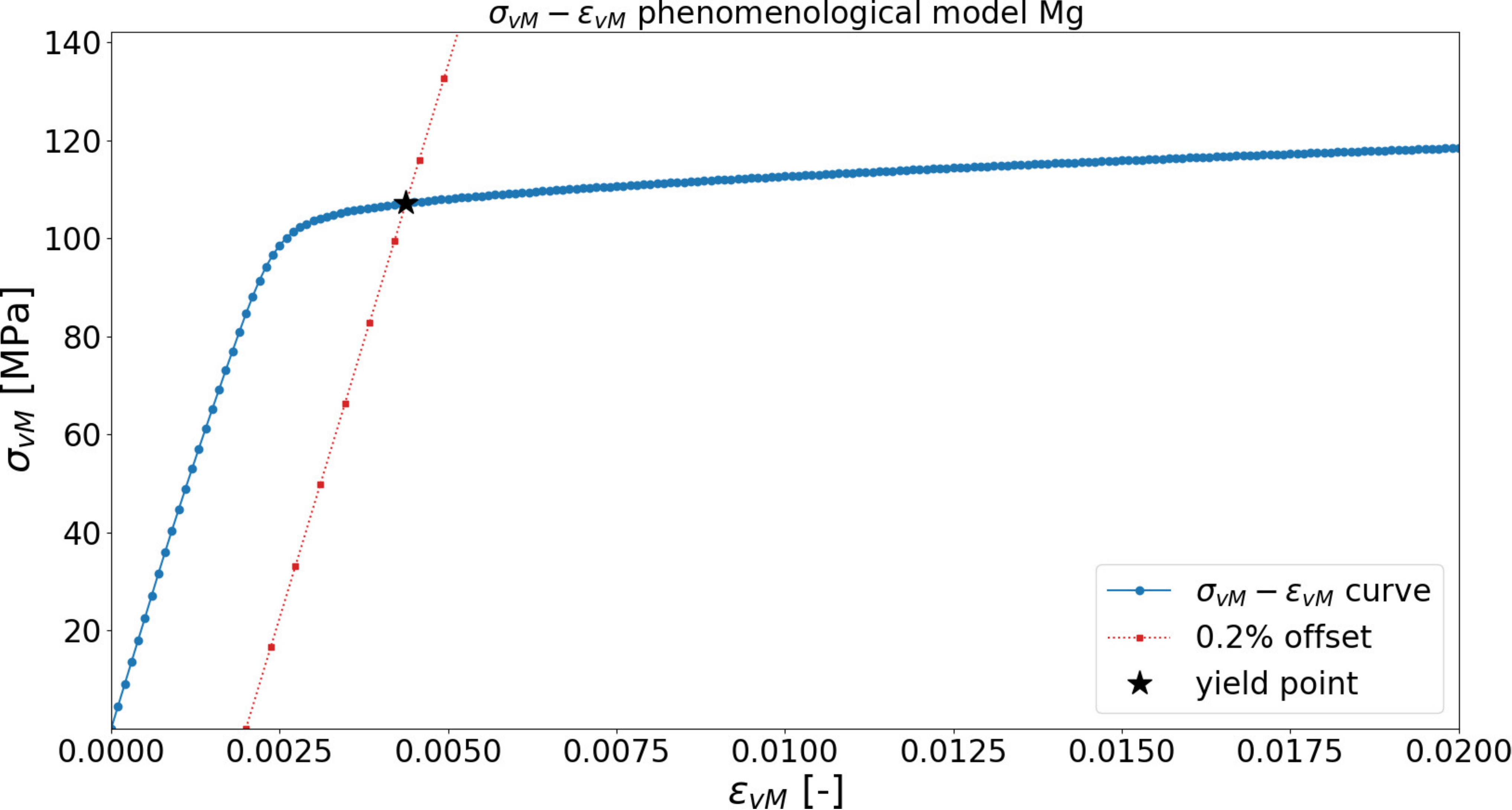}
\caption{Stress-strain equivalent curve for Mg with default parameters and determination of yield point. Modulus of elasticity is estimated as 45.1172 GPa, while $(\varepsilon_\text{Y}, \sigma_\text{Y})$ are estimated as (0.004375, 107.2 MPa)}
\label{fig:cropped_stress-strain-Mg}
\end{figure}

To compare with the default parameter, a single CPFEM simulation is performed with constitutive parameters described in Table~\ref{tab:MgConstitutiveParameters}. 
The stress-strain equivalent curve is shown in Figure~\ref{fig:cropped_stress-strain-Mg}. 
As a reference to experimental data\footnote{\href{https://www.matweb.com/search/DataSheet.aspx?MatGUID=7b49605d472d40d393ffe87ea224980c}{https://www.matweb.com/search/DataSheet.aspx?MatGUID=7b49605d472d40d393ffe87ea224980c}}, 
% Magnesium
% % http://www.matweb.com/search/DataSheet.aspx?MatGUID=7b49605d472d40d393ffe87ea224980c
% Modulus of Elasticity: 44 GPa
% Yield strength: 90-105 MPa
the modulus of elasticity for polycrystalline Mg is reported at 44 GPa, whereas its yield strength is reported as 90-105 MPa. 
Compared to the experiment, the computed modulus and yield strength $\sigma_\text{Y}$ are well calibrated, as the computational results agree very well with the experimental data.

\subsection{Numerical results}

\begin{figure}[!hbtp]
\centering
\includegraphics[width=\textwidth, keepaspectratio]{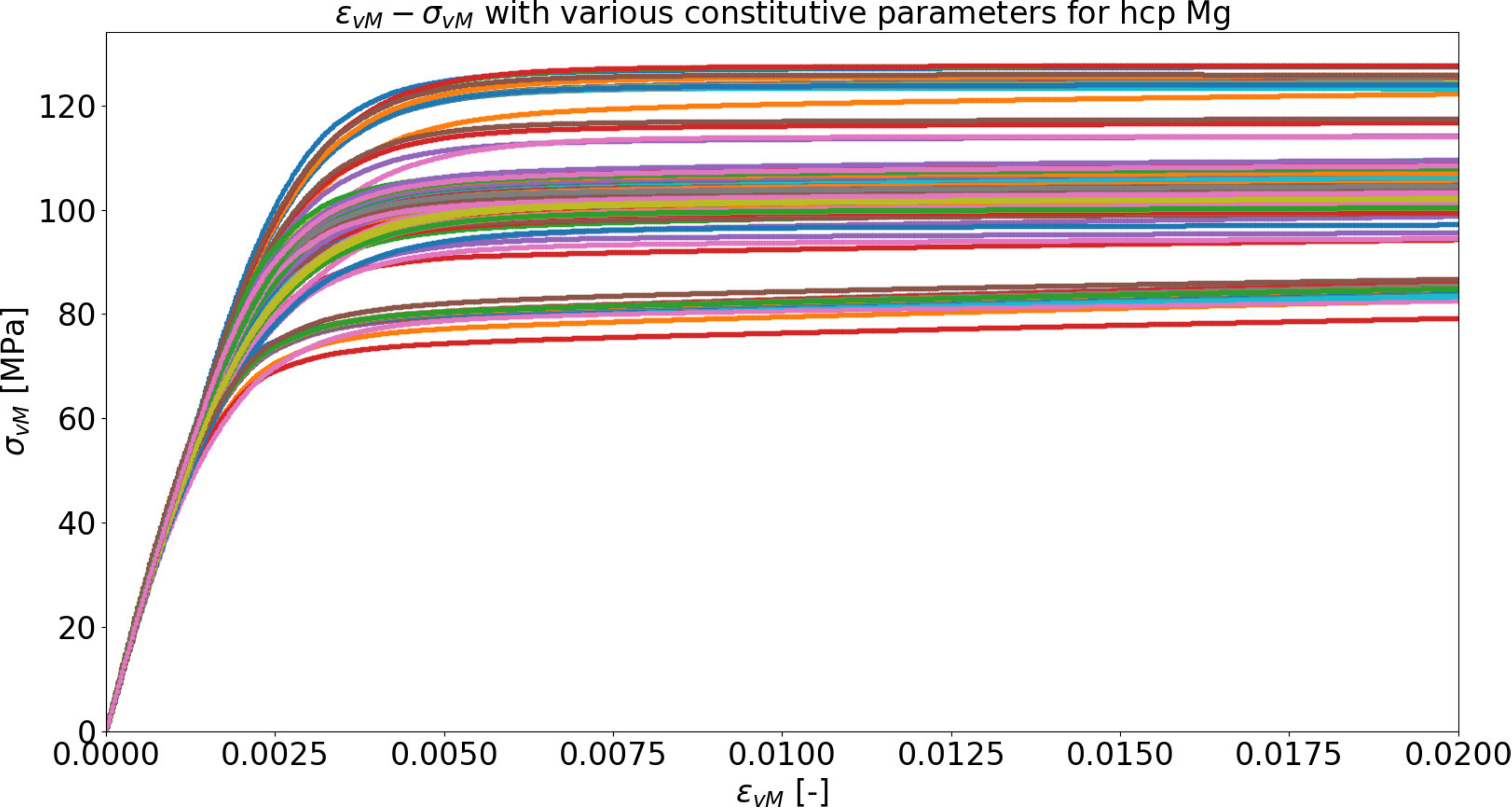}
\caption{Equivalent $\varepsilon_{\text{vM}}-\sigma_{\text{vM}}$ plots for hcp Mg.}
\label{fig:cropped_compiled-stress-strain-Mg-eps-converted-to.pdf}
\end{figure}

Figure~\ref{fig:cropped_compiled-stress-strain-Mg-eps-converted-to.pdf} shows a compilation of stress-strain curves for 577 simulations, where each corresponds to a unique set of constitutive parameters for hcp Mg. As shown in this figure, the constitutive model has a minor effect on the effective modulus of elasticity, and more profound effect on the yield stress $\sigma_{\text{Y}}$.

% \begin{figure}[!htbp]
% \centering
% \includegraphics[width=\textwidth, keepaspectratio]{figsSC-CPFEM/cropped_jointPlot-Mg-eps-converted-to.pdf}
% \caption{Joint probability density plot $\varepsilon_{\text{Y}}-\sigma_{\text{Y}}$ for hcp Mg.}
% \label{fig:cropped_jointPlot-Mg-eps-converted-to.pdf}
% \end{figure}

Figure~\ref{fig:cropped_pdf-strainYield-Mg-eps-converted-to.pdf} and Figure~\ref{fig:cropped_pdf-stressYield-Mg-eps-converted-to.pdf}, respectively, show the probability density function for $\varepsilon_{\text{Y}}$ and $\sigma_{\text{Y}}$. 
The mode for $\varepsilon_{\text{Y}}$ is approximately 0.0054, whereas the mode for $\sigma_{\text{Y}}$ is approximately 99 MPa. 
The uncertainty explained in $\sigma_\text{Y}$ reasonably agree with experimental data.
% It should be noted that with the current sparse grid level $\ell = 3$, the approximation for $\varepsilon_\text{Y}$ is probably not very accurate. 
% One of the possible reasons is that the elastic regime of copper is very small (as shown in Figure~\ref{fig:cropped_compiled-stress-strain-Cu-eps-converted-to.pdf}, and therefore, numerous simulations are required to accurately capture the yield strain and stabilize the polynomial approximation. 

\begin{subfigure}
\setcounter{figure}{12}
\setcounter{subfigure}{0}
  \begin{minipage}{0.475\textwidth}
      \centering  
      \includegraphics[width=\textwidth, keepaspectratio]{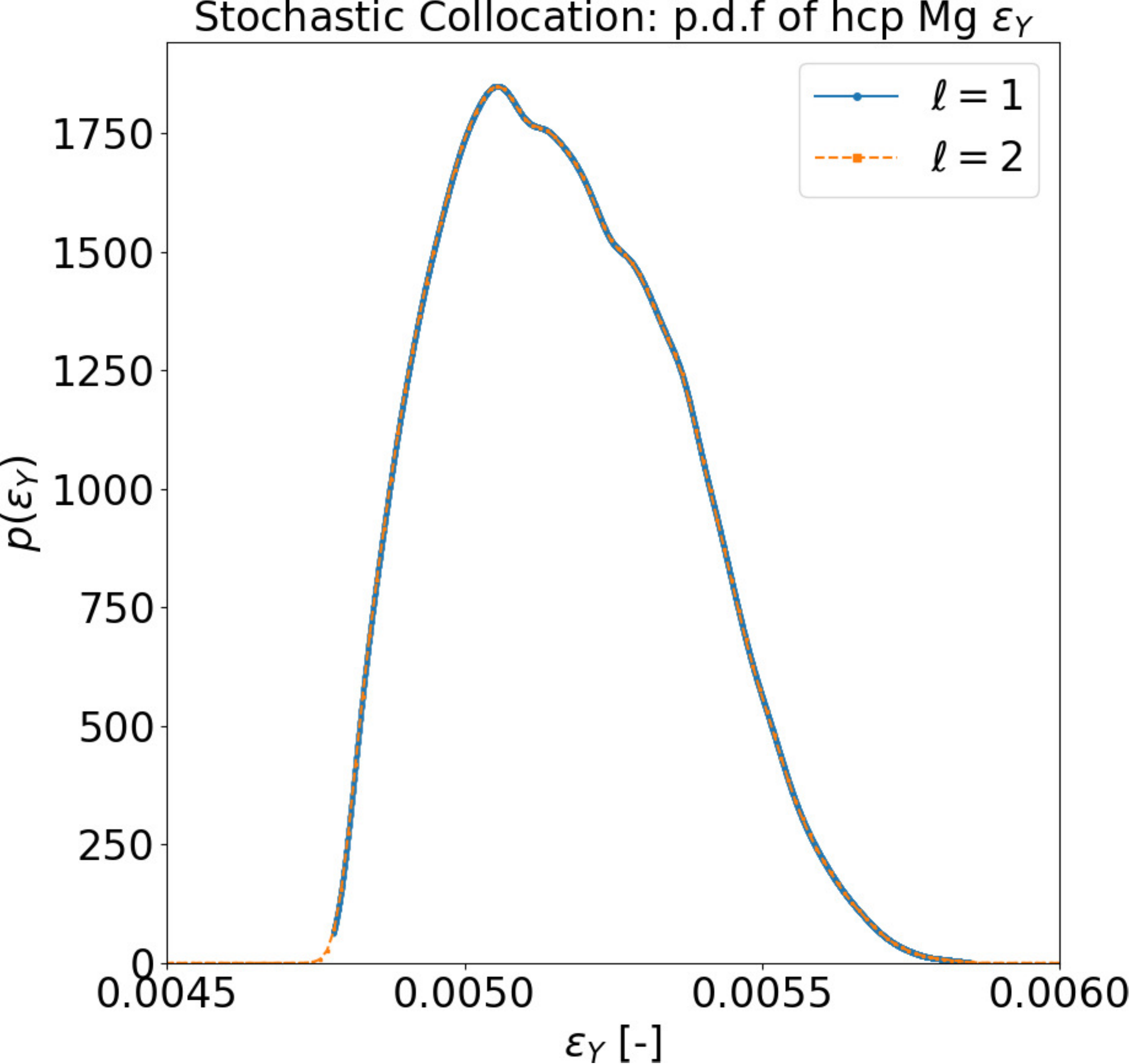}
      \caption{SC probability density function of $\varepsilon_{\text{Y}}$ for hcp Mg.}
      \label{fig:cropped_pdf-strainYield-Mg-eps-converted-to.pdf}
    \end{minipage}\hfill%
    \begin{minipage}{0.475\textwidth}
      \centering
      \includegraphics[width=\textwidth, keepaspectratio]{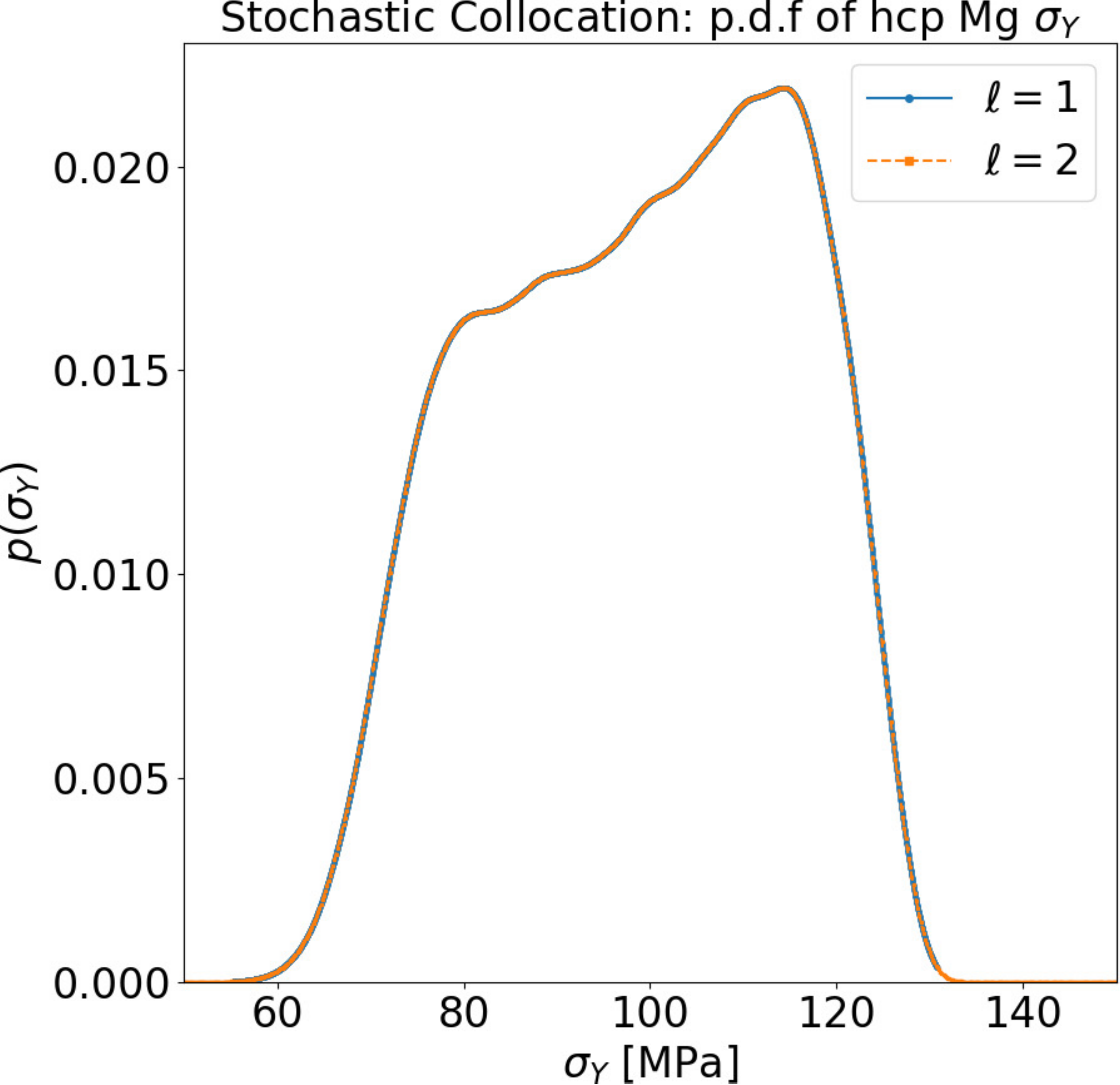}
      \caption{SC probability density function of $\sigma_{\text{Y}}$ for hcp Mg.}
      \label{fig:cropped_pdf-stressYield-Mg-eps-converted-to.pdf}
  \end{minipage}% 
\label{fig:pdf-Mg}
\end{subfigure}

\begin{figure}[!htbp]
\centering
\includegraphics[width=\textwidth, keepaspectratio]{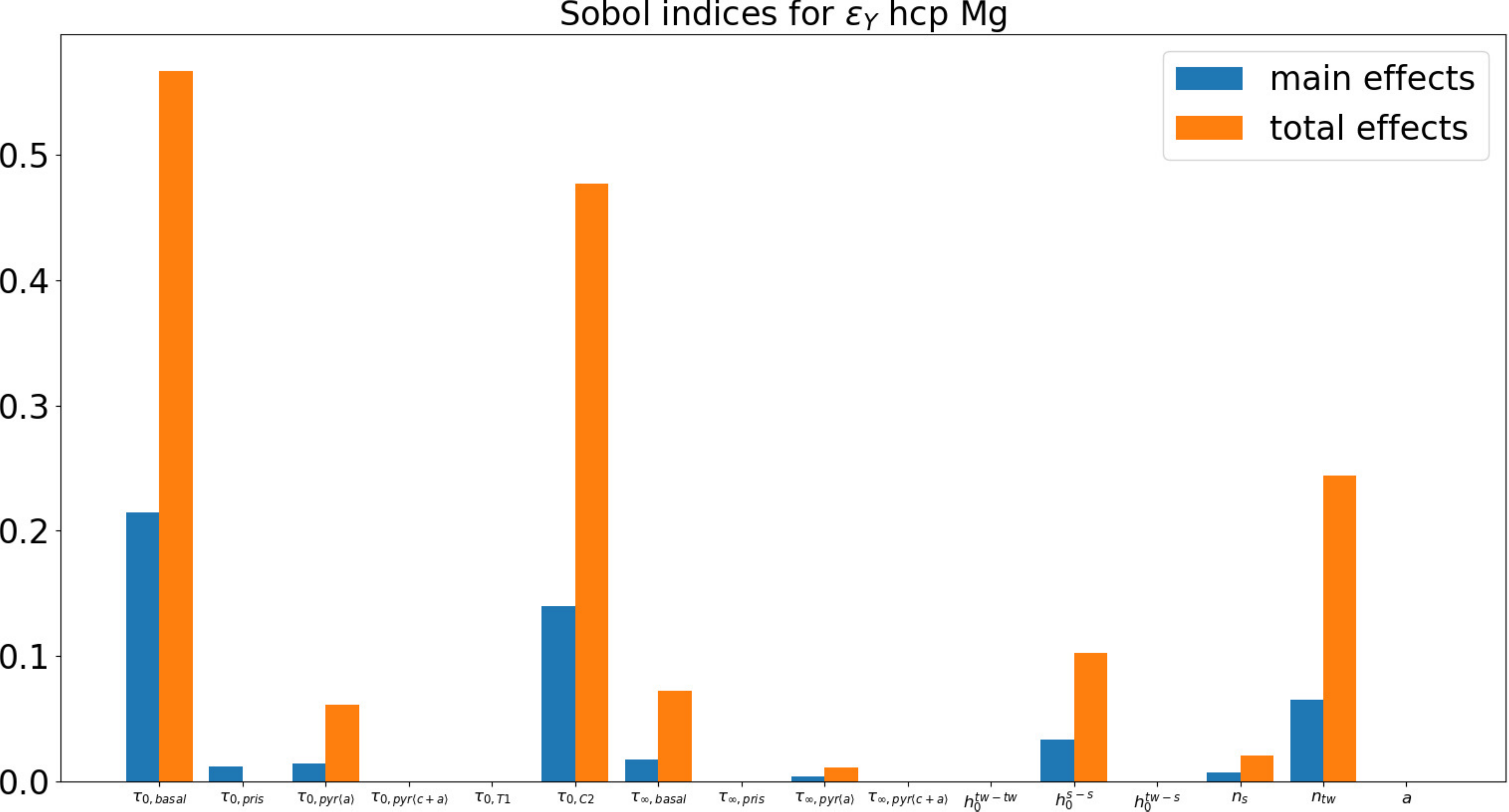}
\caption{Sobol' indices for $\varepsilon_{\text{Y}}$ for hcp Mg.}
\label{fig:cropped_sobol-strainYield-Mg-eps-converted-to.pdf}
\end{figure}

\begin{figure}[!htbp]
\centering
\includegraphics[width=\textwidth, keepaspectratio]{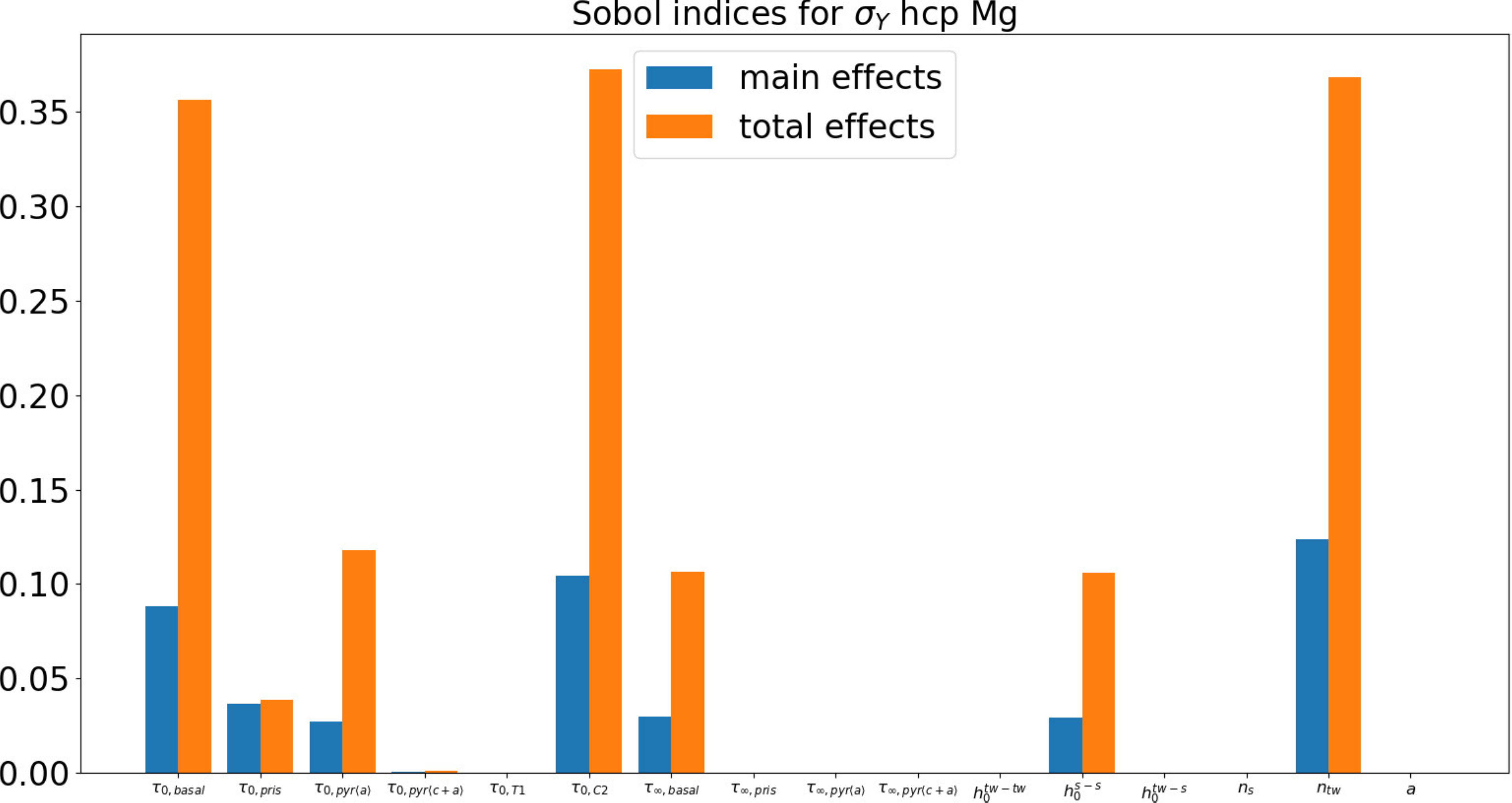}
\caption{Sobol' indices for $\sigma_{\text{Y}}$ for hcp Mg.}
\label{fig:cropped_sobol-stressYield-Mg-eps-converted-to.pdf}
\end{figure}

Figure~\ref{fig:cropped_sobol-strainYield-Mg-eps-converted-to.pdf} and Figure~\ref{fig:cropped_sobol-stressYield-Mg-eps-converted-to.pdf}, respectively, show the Sobol' indices for $\varepsilon_\text{Y}$ and $\sigma_\text{Y}$. 
Ranking from the most influential parameters to the least influential parameters for $\varepsilon_\text{Y}$ from the Sobol indices for main effects, 
$T_{\tau_{0,\text{basal}}} = 0.5668$, 
$T_{\tau_{0,\text{C}2}} = 0.4772$, 
$T_{n_\text{tw}} = 0.2439$, 
$T_{h_{0}^{\text{s}-\text{s}}} = 0.1021$, 
$T_{\tau_{\infty,\text{basal}}} = 0.07249$, 
$T_{\tau_{0,\text{pyr} \langle a \rangle}} = 0.6131$, 
$T_{n_\text{s}} = 0.02082$, 
$T_{\tau_{\infty,\text{pyr} \langle a \rangle}} = 0.01091$.
Ranking from the most influential parameters to the least influential parameters for $\sigma_\text{Y}$ from the Sobol indices for main effects,
$T_{\tau_{0,\text{C}2}} = 0.3729$
$T_{n_\text{tw}} = 0.3684$,
$T_{\tau_{0,\text{basal}}} = 0.3566$,
$T_{\tau_{0,\text{pyr} \langle a \rangle}} = 0.1181$,
$T_{\tau_{\infty,\text{basal}}} = 0.1064$,
$T_{h_{0}^{\text{s}-\text{s}}} = 0.1061$,
$T_{\tau_{0,\text{pris}}} = 0.03861$. 
Compared to Sedighiani et al.~\cite{sedighiani2020efficient,sedighiani2022determination}, our analysis shows some agreements, but mostly differ in the set of sensitive parameters. 
Possible explanations are due to (1) different quantities of interest and (2) methodological approach: Sedighiani et al.~\cite{sedighiani2020efficient,sedighiani2022determination} studies are conducted based on ANOVA, whereas our approach relies on global sensitivity analysis with Sobol' indices.

\section{Dislocation-density-based constitutive model for bcc W}
\label{sec:DisloBccW}

\subsection{Constitutive law}

For the sake of completeness, we adopt the dislocation-density-based constitutive law description from Cereceda et al.~\cite{cereceda2016unraveling,cereceda2015linking}, Stukowski et al.\cite{stukowski2015thermally}, and summarize it here. Interested readers are further referred to Cereceda et al.~\cite{cereceda2016unraveling,cereceda2015linking,cereceda2013assessment}, Stukowski et al.\cite{stukowski2015thermally}, especially~\cite{cereceda2016unraveling} for a complete formulation.

It is assumed that all the plastic deformation is due to dislocation slip, i.e.
\begin{equation}
% \mathbf{L}_p = \sum_a \dot{\gamma}^{\alpha} \mathbf{m}^{\alpha} \otimes \mathbf{n}^\alpha
\mathbf{L}_p = \sum_a \dot{\gamma}^{\alpha} \mathbf{P}^\alpha_\text{S}
\end{equation}
where $\alpha$ is a slip system, $\mathbf{m}^\alpha$ and $\mathbf{n}^\alpha$ are unit vectors in the normalized slip direction and the normal to the slip plane of the system $\alpha$, respectively, $\mathbf{P}^\alpha_\text{S} = \mathbf{m}^{\alpha} \otimes \mathbf{n}^\alpha$ is a Schmid geometric projection tensor. 
The resolved shear stress of slip system $\alpha$ include both Schmid and non-Schmid factors as
\begin{equation}
\tau^{\alpha} = \mathbf{P}^\alpha_{\text{total}} : \mathbf{\sigma} = (\mathbf{P}^\alpha_{\text{S}} + \mathbf{P}^\alpha_{\text{T/AT}} + \mathbf{P}^\alpha_{\text{ng}}) : \mathbf{\sigma}
\end{equation},
where $\mathbf{P}^\alpha_{\text{T/AT}} = a_1 \mathbf{m}^{\alpha} \otimes \mathbf{n}_1^\alpha $ is a non-Schmid tensor representing twinning and anti-twinning asymmetry and the effects due to non-glide stress components, 
$\mathbf{P}^\alpha_{\text{ng}} = a_2 (\mathbf{n}^{\alpha} \otimes \mathbf{m}^\alpha) \otimes \mathbb{n}^{\alpha} + a_3 (\mathbf{n}_1^{\alpha} \otimes \mathbf{m}^\alpha) \otimes \mathbf{n}_1^\alpha $. 
$\mathbf{n}_1^\alpha$ forms an angle of $-60^\circ$ with the reference slip plane defined by $\mathbf{n}^\alpha$ and changes sign with the direction of slip on each glide plane~\cite{koester2012atomistically}. 
$a_1, a_2, a_3$ are calibrated material-dependent constants.
The shear rate $\dot{\gamma}^\alpha$ is given by the Orowan equation
\begin{equation}
% \dot{\gamma}^\alpha = \rho^{\alpha} b v^{\alpha},
% \dot{\gamma}^\alpha = b \rho^{\alpha} v^{\alpha},
\dot{\gamma}^\alpha = b \rho^{\alpha} v_s(\tau^{\alpha}, T),
\end{equation}
where $b = a_0 \sqrt{3} / 2$ is the magnitude of the Burgers vector, $a_0$ is the lattice parameter, $T$ is the absolute temperature, $\rho^\alpha$ is the density of mobile screw dislocations in slip system $\alpha$, and $v_s(\tau^\alpha, T)$ is the screw dislocation velocity, which captures the thermally activated character of dislocation motion. 

Under the assumption that kink relaxation is significantly faster than kink-pair nucleation, the total time $t_t$ required for a kink pair to form and sweep a rectilinear screw dislocation segment of length $\lambda^\alpha$ is
\begin{equation}
t_t = t_n + t_k = J(\tau^\alpha, T)^{-1} + \frac{\lambda^\alpha - w}{2 v_k(\tau^\alpha, T)},
\end{equation}
where $t_n$ is the mean time to nucleate a kink pair, $t_k$ is the time needed for a kink to sweep half a segment length, $J$ is the kink-pair nucleation rate, $w$ is the kink-pair separation, $v_k$ is the kink velocity.

The kink-pair nucleation rate is modeled by an Arrhenius formulation as
\begin{equation}
J(\tau^\alpha, T) = \frac{v_0 (\lambda^\alpha - w)}{b} \exp\left( -\frac{\Delta H_{\text{kp}}(\tau^\alpha)}{kT} \right),
\end{equation}
where $v_0$ is an attempt frequency, $\Delta H_{\text{kp}}$ is the activation enthalpy of a kink pair stress $\tau^\alpha$, $k$ is Boltzmann's constant. 
The kink velocity is modeled as
\begin{equation}
v_k(\tau^\alpha, T) = \frac{b \tau^\alpha}{B(T)},
\end{equation}
where $B$ is simplified to a constant. The dislocation velocity can be modeled as
\begin{equation}
v_s = \frac{h}{t_t} = \frac{h}{t_n + t_k} = \frac{2 b h \tau^\alpha v_0 (\lambda^\alpha - w) \exp \left( -\frac{\Delta H_{\text{kp}}}{kT} \right)}{2 b^2 \tau^\alpha + v_0 B (\lambda^\alpha - w)^2 \exp \left( -\frac{\Delta H_{\text{kp}}}{kT} \right) },
\end{equation}
where $h= a_0 \sqrt{6}/3$ is the distance between two consecutive Peierls valleys. When $t_k \ll t_n $, $v_0 B (\lambda^\alpha - w)^2 \exp \left( -\frac{\Delta H_{\text{kp}}}{kT} \right) \to 0$, and the common diffusive velocity expression is recovered
\begin{equation}
v_s  = v_0 h \frac{(\lambda^\alpha - w)}{b} \exp \left( - \frac{\Delta H_{\text{kp}}(\tau^\alpha)}{kT} \right) \text{sgn}(\tau^\alpha).
\end{equation}
It is further elaborated in ~\cite{sedighiani2020efficient} that 
\begin{equation}
\Delta H_{\text{kp}}(\tau^\alpha) = \Delta H_{\text{kp}} \left[  1 - \left(\frac{\tau_T^{\alpha}}{\tau_0^*}\right)^p \right]^q,
\end{equation}
where $p$ and $q$ determine the shape of the short-range activation energy. 

% \begin{equation}
% \tau^\chi_C = \frac{\tau_{\text{Peierls}} + \sigma \left(a_2 \sin (2\chi) + a_3 (2 \chi + \frac{\pi}{6}) \right) }{ \cos \chi + a_1 \cos \left( \chi + \frac{\pi}{3} \right) }
% \end{equation}

Following the Kocks-Mecking family of dislocation density evolution models~\cite{mecking1981kinetics}, the mobile dislocation density on slip system $\alpha$ evolves in time is modeled as
\begin{equation}
\dot{\rho}^\alpha = \dot{\rho}^\alpha_\text{mult} + \dot{\rho}^\alpha_\text{ann},
\end{equation}
with the initial dislocation density $\rho^\alpha (t=0) = \rho^\alpha_0$. 
Dislocation multiplication is proportional to the inverse mean free path of the dislocation $\lambda^\alpha$ and the plastic strain rate, as
\begin{equation}
\dot{\rho}^\alpha_{\text{mult}} =  \frac{\dot{\gamma}^\alpha}{b\lambda^\alpha}.
\end{equation}
$\lambda^\alpha$ is defined as
\begin{equation}
\frac{1}{\lambda^\alpha} = \frac{1}{d_g} + \frac{\sqrt{\rho^\alpha_f}}{c},
\end{equation}
where $d_g$ is the grain size, $c$ is a hardening constant, $\rho^\alpha_f$ is the forest dislocation density and calculated as
\begin{equation}
\rho^\alpha_f = \sum_\beta \rho^\beta | \mathbf{n}^\alpha \cdot \mathbf{m}^\beta |
\end{equation}

Dislocation annihilation occurs spontaneously when dipoles approach within a spacing of $d_\text{edge}$
\begin{equation}
\dot{\rho}^\alpha_\text{ann} = - \frac{2 d_\text{edge}}{b} \rho^\alpha \left| \dot{\gamma}^\alpha \right|.
\end{equation}
The resolved shear stress $\tau^\alpha$ is corrected by replacing with $\tau^{\alpha'}$ and accounting for the latent and self-hardening as
\begin{equation}
\tau^{\alpha'} = \tau^\alpha - \tau_\text{h} = \mathbf{P}_\text{tot}^\alpha : \mathbf{\sigma} - \mu b \sqrt{\sum_{\alpha'} \xi_{\alpha \alpha'} \rho^{\alpha'}},
\end{equation}
where $\tau_\text{h}$ is the hardening stress, $\xi_{\alpha \alpha'}$ are the coefficients of the interaction matrix between slip system $\alpha$ and $\alpha'$, respectively, as six possible independent interactions: self, coplanar, collinear, orthogonal, glissile, and sessile. 

\begin{table}[!htbp]
\tiny
\caption{Parameters for W used in this case study (cf. Tables 2 and 3 in Cereceda et al.~\cite{cereceda2016unraveling,cereceda2015linking}, Section 6.2.4~\cite{roters2019damask}). Bounds for $\tau_{\text{Peierls}}$ are devised based on Samolyuk et al~\cite{samolyuk2012influence} and Cereceda et al~\cite{cereceda2016unraveling}.}
\label{tab:WConstitutiveParameters}
\begin{tabular*}{\textwidth}{c @{\extracolsep{\fill}} cccccc} \hline
variable                  & description                             & units     &  reference value      &  nature       &  distribution          \\ \hline
$C_{11}$                  & elastic constant                        &   GPa     &  523.0          &  deterministic  &  --             \\   
$C_{12}$                  & elastic constant                        &   GPa     &  202.0          &  deterministic  &  --             \\
$C_{44}$                  & elastic constant                        &   GPa     &  161.0          &  deterministic  &  --             \\
$b$                       & Burgers vector                          &  nm       &  $0.272        $  &  deterministic  &  --         \\  \hline
$\nu_0    $               & initial dislocation glide velocity      &  m/s      &  $1.0  \times 10^{-4}$  &  stochastic     &  $\mathcal{U}[10^{-3},\cdot 10^{-5}]$         \\   
$\rho_0^\alpha$           & initial dislocation density             &  m/m$^3$  &  $1.0  \times 10^{12}$  &  stochastic     &  $\mathcal{U}[10^{10},5 \cdot 10^{12}]$   \\   
$\tau_{\text{Peierls}}$   & Peierls stress                          &  GPa      &  $2.03$           &  stochastic     &  $\mathcal{U}[1.64,2.42] $   \\
$p$                       & $p$-exponent in glide velocity, $0 < p \leq 1$            &  --       &  $0.32         $  &  stochastic     &  $\mathcal{U}[0.1,1.0]$         \\   
$q$                       & $q$-exponent in glide velocity, $1 \leq q \leq 2$            &  --       &  $1.46         $  &  stochastic     &  $\mathcal{U}[1.0,2.0]$         \\   
$\Delta H_0   $           & activation energy for dislocation glide &  J        &  $2.725\times 10^{-19}$   &  stochastic     &  $\mathcal{U}[1.5 , 3.5 ] \cdot 10^{-19}$         \\ 
% $\tau^*_0$       & short-range barriers strength at 0K         &  MPa    &  $395        $  &  stochastic     &  $\mathcal{U}[250,500]$         \\   
% $B$          & drag coefficient                  &  Pa s     &  $      10^{-4}$  &  stochastic     &  $\mathcal{U}[10^{-2},10^{-5}]$         \\   
% $d_g$          & average grain size                &  $\mu$ m  &  $50         $  &  stochastic     &  $\mathcal{U}[30, 100]$         \\   
$C_{\lambda}$             & dislocation mean free path parameter    &  --       &  $10         $  &  stochastic     &  $\mathcal{U}[ 5,20]$         \\   \hline
% $C_{\text{anni}}$    &  dislocation annihilation coefficient       &  --     &  $8.7        $  &  stochastic     &  $\mathcal{U}[5,25]$         \\   

\end{tabular*}
\normalsize
\end{table}
 
% \chhighlight{I CANNOT FIND p and q IN EQUATIONS.}

\subsection{Design of numerical experiments}

For dislocation-density-based CPFEM simulations, we used an RVE shown in Figure~\ref{fig:tungstenDislocationDensityBasedRVE}, with the crystallographic texture shown in Figure~\ref{fig:rotatedCubeComponentODF} and Euler angles $(\phi_1, \theta, \phi_2) = (0,0,45)$. Average grain size of 3.0339$\mu$m is used, where an RVE of size 32$\mu$m$^3$ is generated. A mesh of 32$^3$ is created to approximate the microstructure RVE. 
In DREAM.3D, the texture crystallography with Euler angles $(\phi_1, \theta, \phi_2) = (0^\circ,0^\circ,45^\circ)$ is used, the grain size parameters are set as $\mu_D = 1.0986$, $\sigma_D = 0.15$, which results in a RVE with 2089 grains, shown in Figure~\ref{fig:tungstenDislocationDensityBasedRVE}. 
For DAKOTA, we set the sparse grid level $\ell=3$, dimensionality $n=17$, which results in 799 inputs. 
For each set of input parameters, a CPFEM simulation is performed, followed by the post-process. The results are analyzed in the following section. 
A uniaxial loading condition is applied in the [100] direction with $\dot{\varepsilon} = 10^{-3} \text{s}^{-1}$.

\begin{subfigure}
\setcounter{figure}{15}
\setcounter{subfigure}{0}
  \begin{minipage}{0.25\textwidth}
      \centering  
      \includegraphics[width=\linewidth]{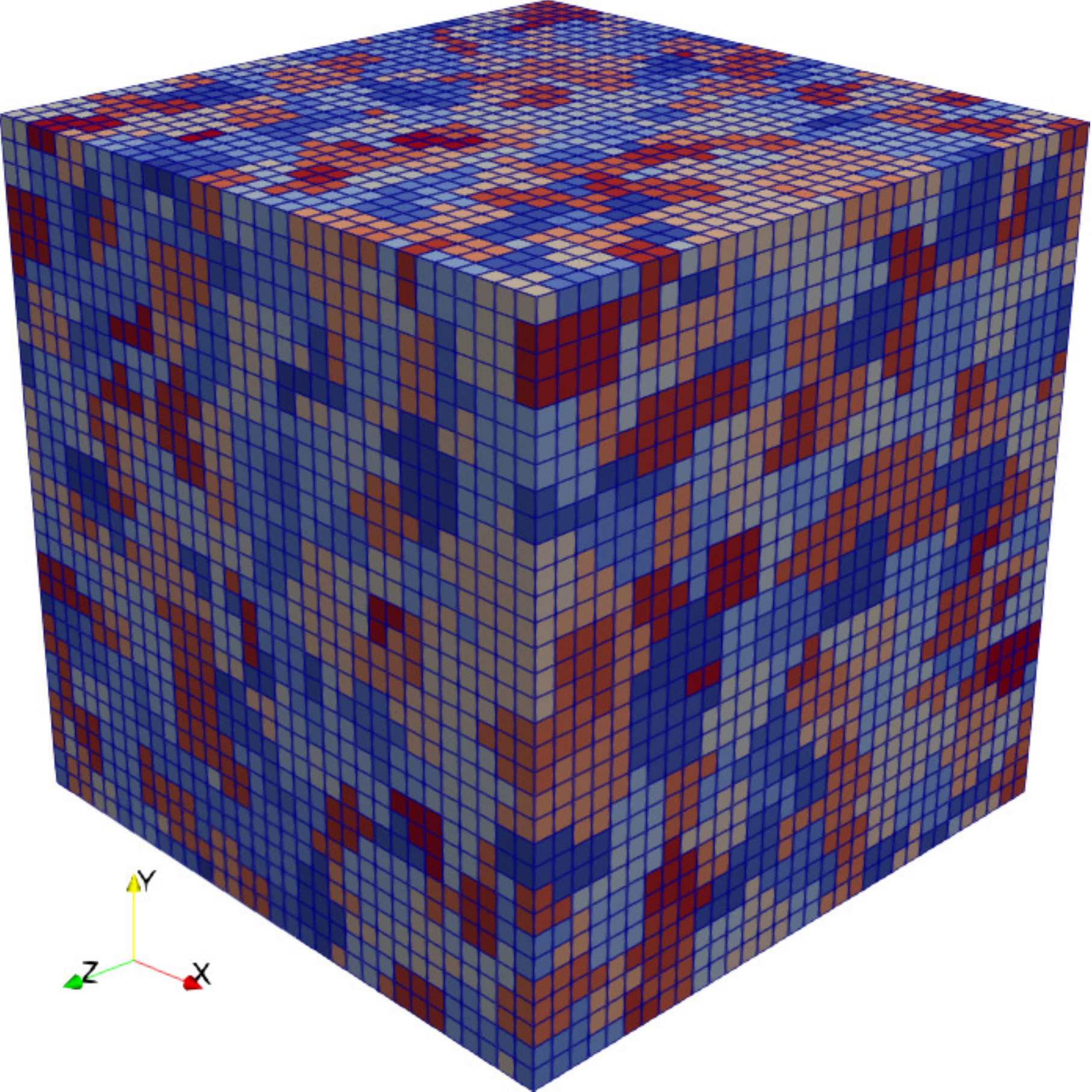}
      \caption{Representative volume element for W.}
      \label{fig:tungstenDislocationDensityBasedRVE}
    \end{minipage}\hfill%
    \begin{minipage}{0.70\textwidth}
      \centering
      \includegraphics[width=\linewidth]{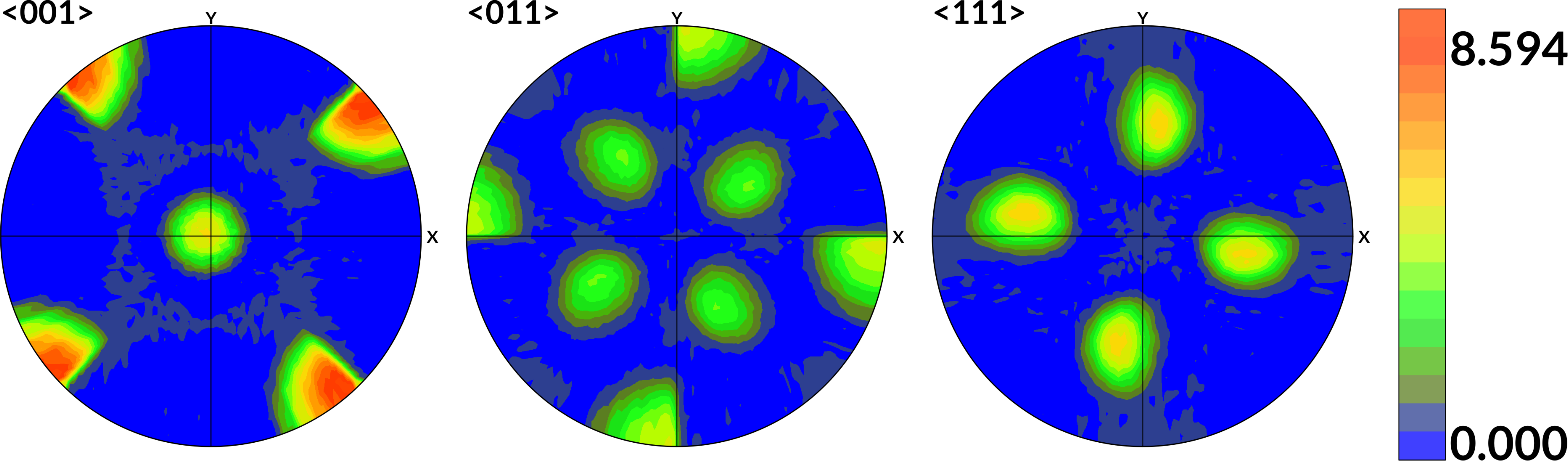}
      \caption{Rotated Cube type of texture component with Euler angles $(\phi_1, \theta, \phi_2) = (0^\circ,0^\circ,45^\circ)$.}
      \label{fig:rotatedCubeComponentODF}
  \end{minipage}% 
% \caption{Microstructure volume element and texture crystallography used in the phenomenological case study.}
\label{fig:dislocationdensitybasedMs}
\end{subfigure}

\begin{figure}
\centering
\includegraphics[width=\textwidth]{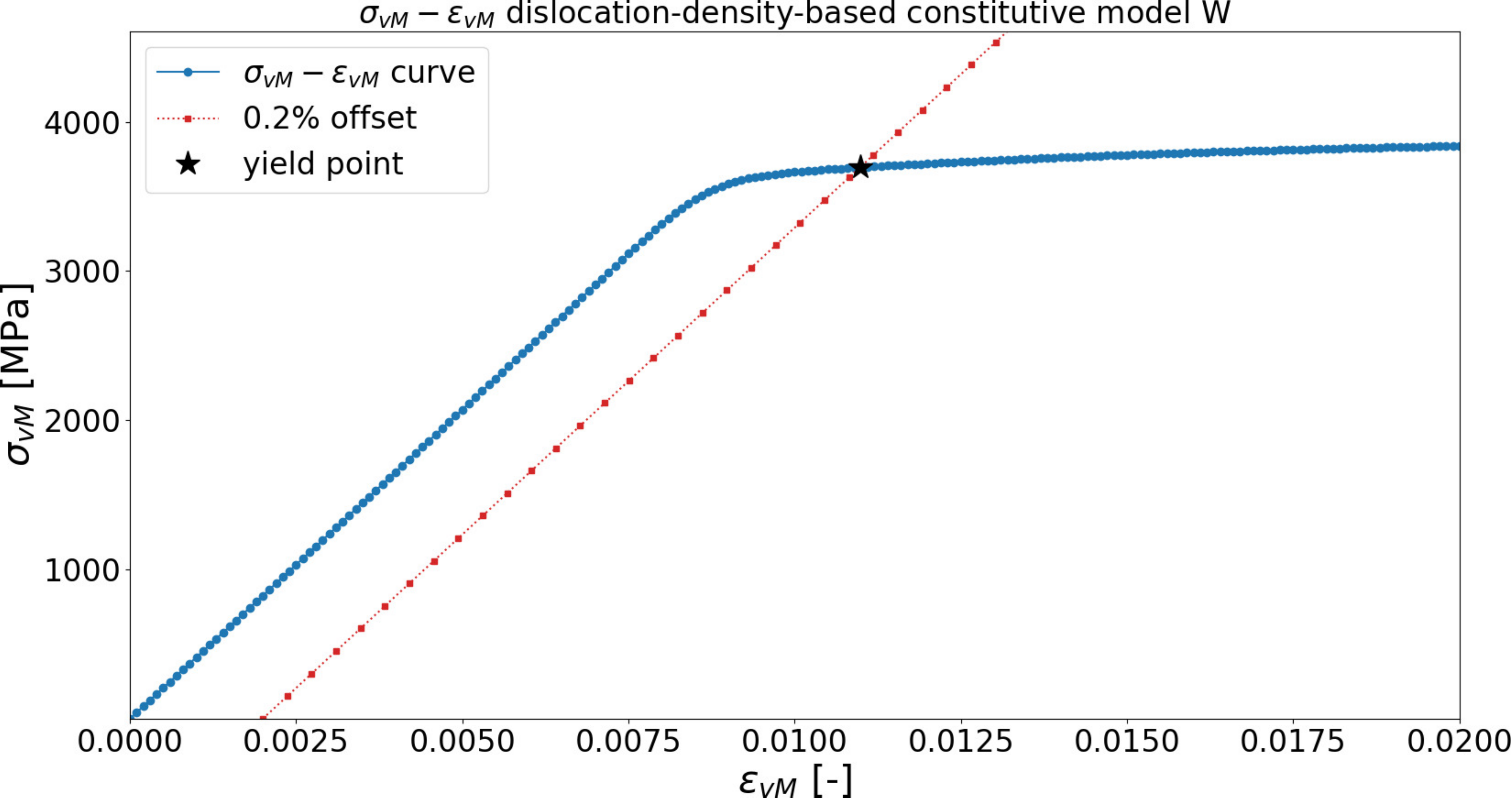}
\caption{Stress-strain equivalent curve for W with default parameters and determination of yield point. Modulus of elasticity is estimated as 411.3528 GPa, while $(\varepsilon_\text{Y}, \sigma_\text{Y})$ are estimated as (0.01098, 3695 MPa).}
\label{fig:cropped_stress-strain-W}
\end{figure}

To compare with the default parameter, a single CPFEM simulation is performed with constitutive parameters described in Table~\ref{tab:WConstitutiveParameters}. 
The stress-strain equivalent curve is shown in Figure~\ref{fig:cropped_stress-strain-W}. 
As a reference to experimental data\footnote{\href{http://www.matweb.com/search/DataSheet.aspx?MatGUID=41e0851d2f3c417ba69ea0188fa570e3}{http://www.matweb.com/search/DataSheet.aspx?MatGUID=41e0851d2f3c417ba69ea0188fa570e3}},
% Tungsten
% % http://www.matweb.com/search/DataSheet.aspx?MatGUID=41e0851d2f3c417ba69ea0188fa570e3
% Modulus of Elasticity: 400 GPa
% Yield strength: 750 MPa
the modulus of elasticity for polycrystalline W is reported at 400 GPa, whereas its yield strength is reported as approximately 750 MPa. 
Compared to the experiment, the computed modulus agrees very well, but the yield strength $\sigma_\text{Y}$ differs, possibly due to the constitutive model is calibrated for single crystal W, whereas the experimental data is reported for polycrystalline W. 
% \chhighlight{I AM NOT SO CONVINCED HERE - IT IS DIFFICULT TO IMAGINE WHY YOU WOULD NEED TO USE MUCH SOFTER MATERIAL PROPERTIES FOR POLYCRYSTALS. ALSO, IF THIS IS THE CASE, DON'T WE NEED TO CALIBRATE PARAMTERS FROM POLYCRYSTALS? }

% \begin{figure}[ht]
% \small
% \begin{minipage}{0.35\textwidth}
% \centering
% \includegraphics[height=80px,keepaspectratio]{figsSC-CPFEM/cropped_tungsten_rve-eps-converted-to.pdf}
% 
% (a) caption here
% \end{minipage}%
% \begin{minipage}{0.60\textwidth}
% \centering
% \includegraphics[height=80px,keepaspectratio]{figsSC-CPFEM/cropped_tungstenODF-eps-converted-to.pdf}
% 
% (b) caption here
% \end{minipage}
% \caption{bla bla bla}
% \label{fig:test}%
% \end{figure}

% \begin{figure}[ht]
% \small
% \begin{minipage}{0.5\textwidth}
% \centering
% \includegraphics[width=0.6\textwidth]{example-image-duck}   \\%{1.jpeg}}
% (a) caption here
% \end{minipage}%
% \begin{minipage}{0.5\textwidth}
% \centering
% \includegraphics[width=0.6\textwidth]{example-image-duck}   \\%{2.jpeg}}
% (b) caption here
% \end{minipage}
% \caption{bla bla bla}
% \label{fig:test}%
% \end{figure}

\subsection{Numerical results}

\begin{figure}[!hbtp]
\centering
\includegraphics[width=\textwidth, keepaspectratio]{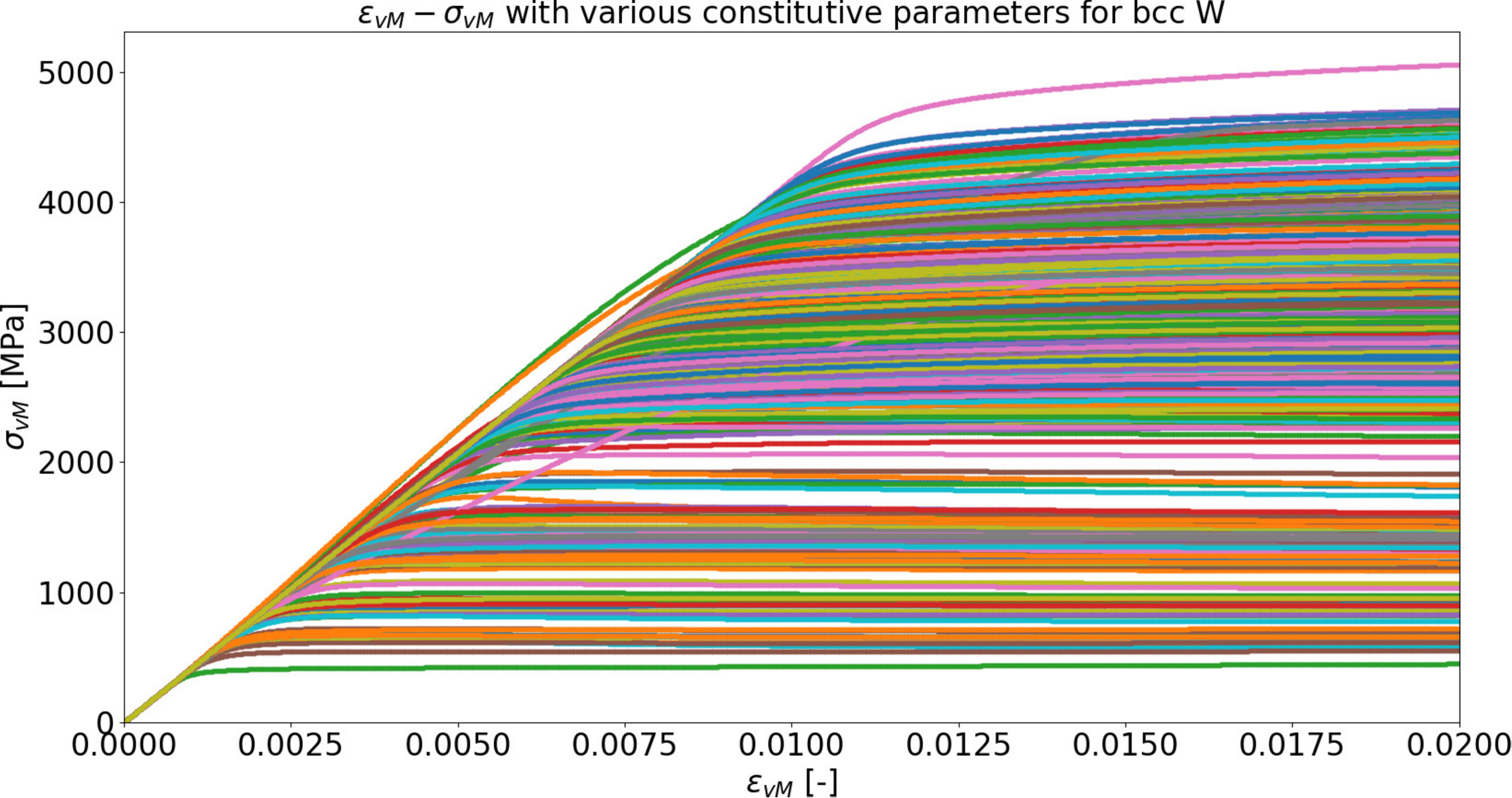}
\caption{Equivalent $\varepsilon_{\text{vM}}-\sigma_{\text{vM}}$ plots for bcc W.}
\label{fig:cropped_compiled-stress-strain-W-eps-converted-to.pdf}
\end{figure}

Figure~\ref{fig:cropped_compiled-stress-strain-W-eps-converted-to.pdf} shows a compilation of stress-strain curves for 799 simulations, where each corresponds to a unique set of constitutive parameters for bcc W. As shown in this figure, the constitutive model has a minor effect on the effective modulus of elasticity, and more profound effect on the yield stress $\sigma_{\text{Y}}$.

% \begin{figure}[!htbp]
% \centering
% \includegraphics[width=\textwidth, keepaspectratio]{figsSC-CPFEM/cropped_jointPlot-W-eps-converted-to.pdf}
% \caption{Joint probability density plot $\varepsilon_{\text{Y}}-\sigma_{\text{Y}}$ for bcc W.}
% \label{fig:cropped_jointPlot-W-eps-converted-to.pdf}
% \end{figure}

Figures~\ref{fig:cropped_pdf-strainYield-W-eps-converted-to.pdf} and Figure~\ref{fig:cropped_pdf-stressYield-W-eps-converted-to.pdf} shows the probability density function for $\varepsilon_{\text{Y}}$ and $\sigma_{\text{Y}}$, respectively. 
The mode for $\varepsilon_{\text{Y}}$ is approximately 0.01010, whereas the mode for $\sigma_{\text{Y}}$ is approximately 3350 MPa. 
The uncertainty explained in $\sigma_\text{Y}$ reasonably agree with experimental data. 
For experimental data with $\sigma_\text{Y} \approx 750 $ MPa, fairly a few set of parameters can reproduce and calibrate accordingly, as shown in Figure~\ref{fig:cropped_compiled-stress-strain-W-eps-converted-to.pdf}. 
% It should be noted that with the current sparse grid level $\ell = 3$, the approximation for $\varepsilon_\text{Y}$ is probably not very accurate. 
% One of the possible reasons is that the elastic regime of copper is very small (as shown in Figure~\ref{fig:cropped_compiled-stress-strain-Cu-eps-converted-to.pdf}, and therefore, numerous simulations are required to accurately capture the yield strain and stabilize the polynomial approximation. 

\begin{subfigure}
\setcounter{figure}{18}
\setcounter{subfigure}{0}
  \begin{minipage}{0.475\textwidth}
      \centering  
      \includegraphics[width=\textwidth, keepaspectratio]{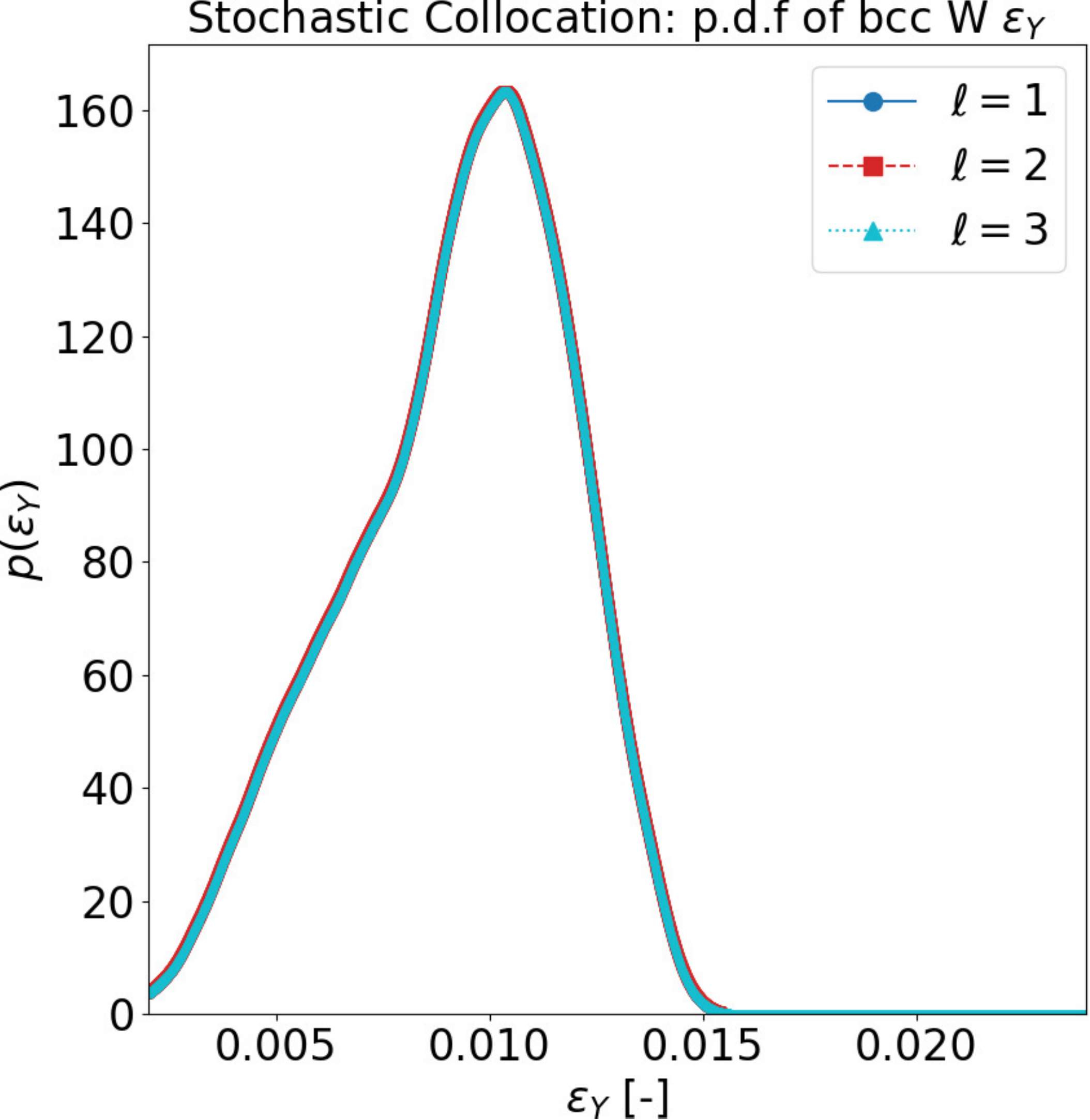}
      \caption{SC probability density function of $\varepsilon_{\text{Y}}$ for bcc W.}
      \label{fig:cropped_pdf-strainYield-W-eps-converted-to.pdf}
    \end{minipage}\hfill%
    \begin{minipage}{0.475\textwidth}
      \centering
      \includegraphics[width=\textwidth, keepaspectratio]{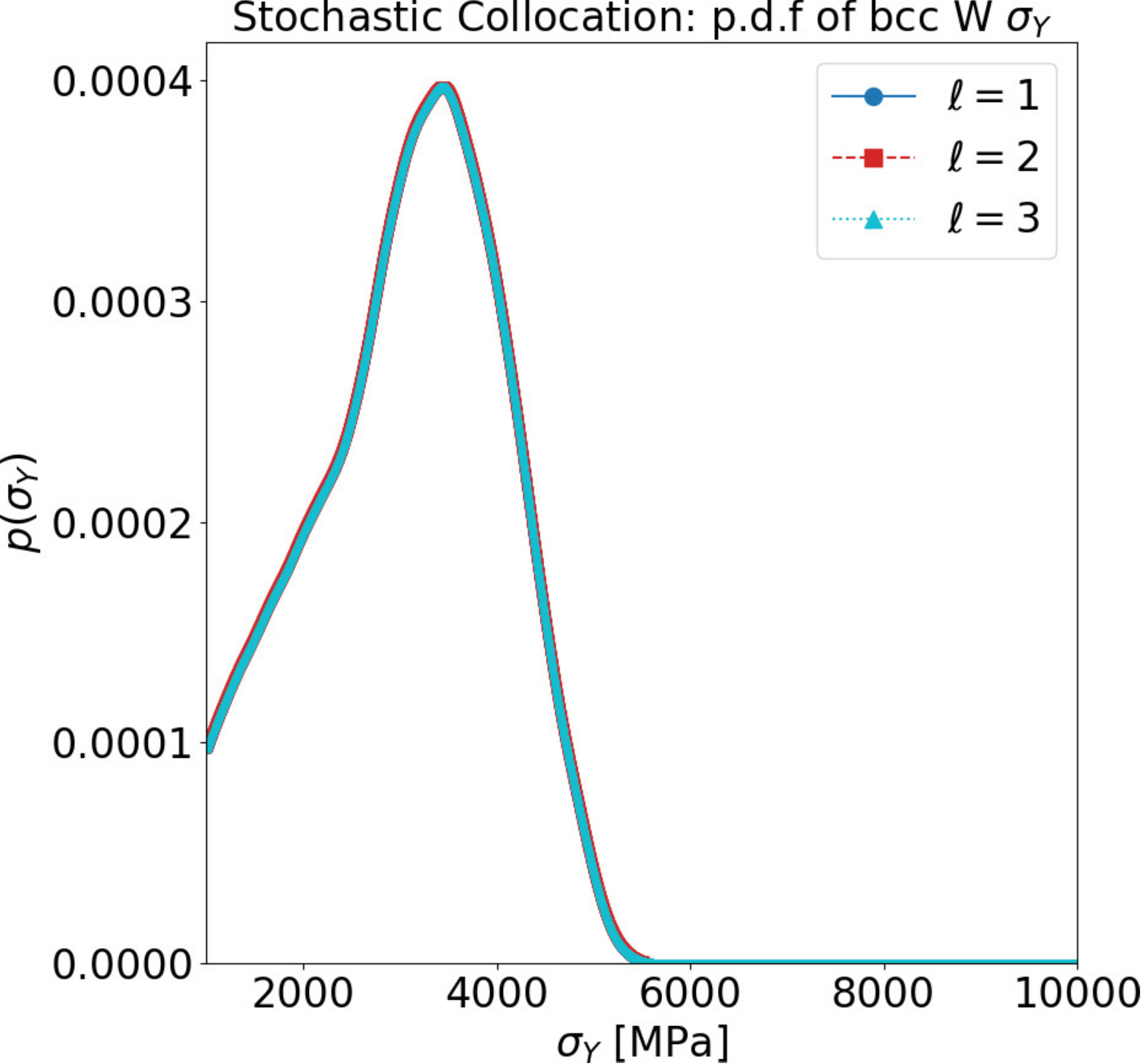}
      \caption{SC probability density function of $\sigma_{\text{Y}}$ for bcc W.}
      \label{fig:cropped_pdf-stressYield-W-eps-converted-to.pdf}
  \end{minipage}% 
\label{fig:pdf-W}
\end{subfigure}

\begin{figure}[!htbp]
\centering
\includegraphics[width=\textwidth, keepaspectratio]{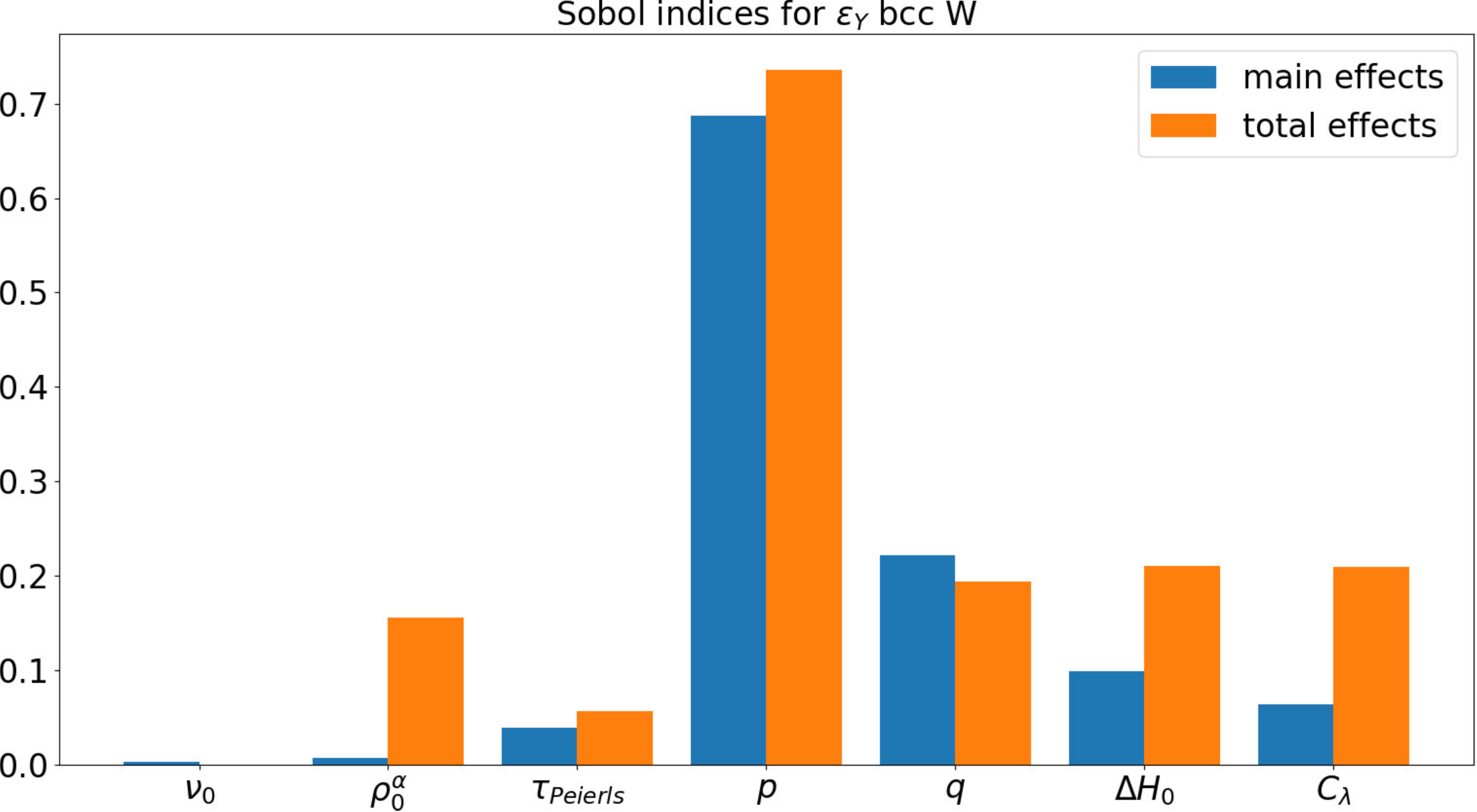}
\caption{Sobol' indices for $\varepsilon_{\text{Y}}$ for bcc W.}
\label{fig:cropped_sobol-strainYield-W-eps-converted-to.pdf}
\end{figure}

\begin{figure}[!htbp]
\centering
\includegraphics[width=\textwidth, keepaspectratio]{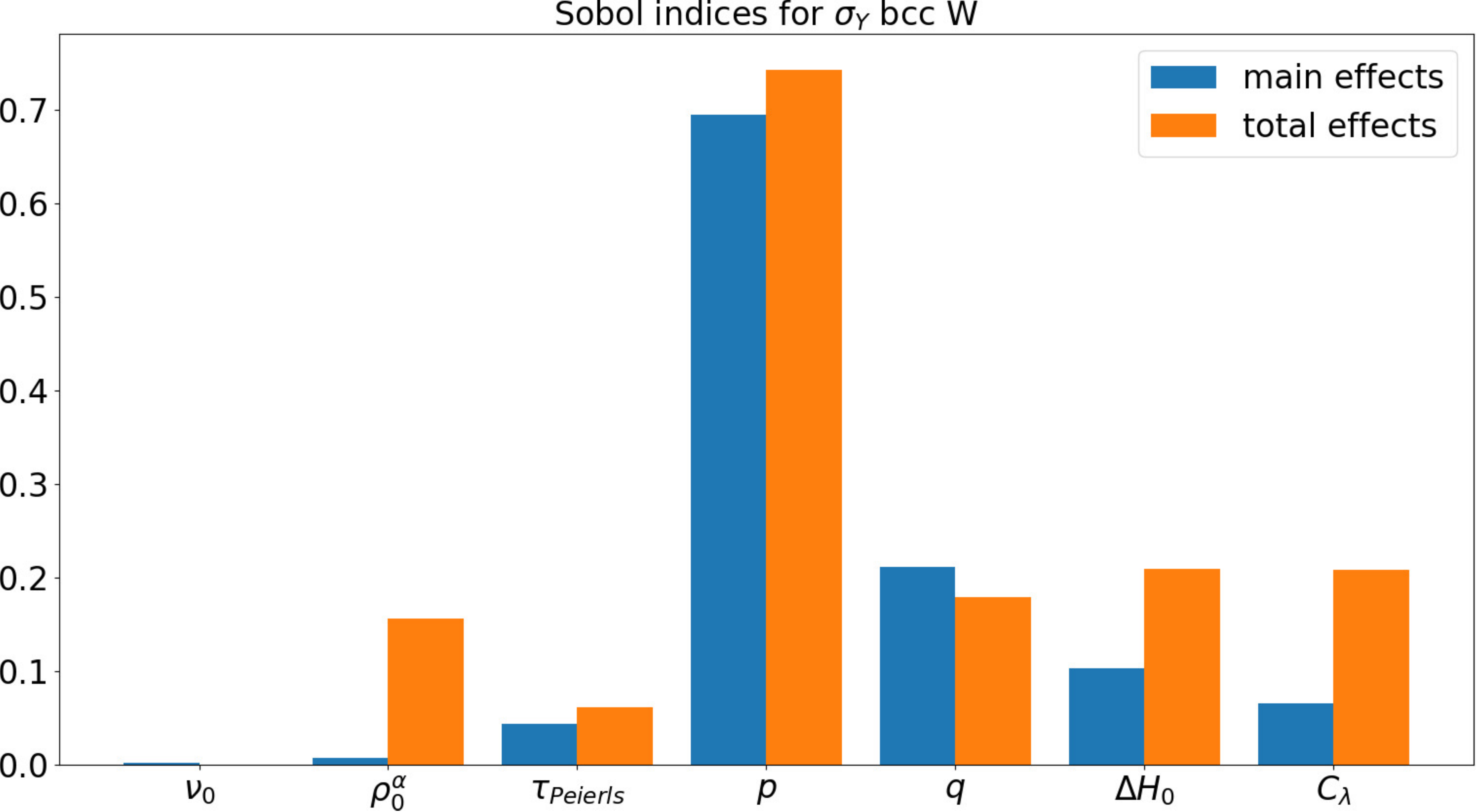}
\caption{Sobol' indices for $\sigma_{\text{Y}}$ for bcc W.}
\label{fig:cropped_sobol-stressYield-W-eps-converted-to.pdf}
\end{figure}

Figure~\ref{fig:cropped_sobol-strainYield-W-eps-converted-to.pdf} and Figure~\ref{fig:cropped_sobol-stressYield-W-eps-converted-to.pdf}, respectively, show the Sobol' indices for $\varepsilon_\text{Y}$ and $\sigma_\text{Y}$. 
Ranking from the most influential parameters to the least influential parameters for $\varepsilon_\text{Y}$ from the Sobol indices for main effects, 
$T_{p} = 0.7356$, 
$T_{\Delta H_0} = 0.212$, 
$T_{C_{\lambda}} = 0.2089$, 
$T_{q} = 0.1942$, 
$T_{\rho_0^\alpha} = 0.1556$, 
$T_{\tau_{\text{Peierls}}} = 0.05712$, 
$T_{\nu_0} = -0.03444$. 
Ranking from the most influential parameters to the least influential parameters for $\sigma_\text{Y}$ from the Sobol indices for main effects, 
$T_{p} = 0.5668$, 
$T_{\Delta H_0} = 0.2095$, 
$T_{C_{\lambda}} = 0.2079$, 
$T_{q} = 0.1795$, 
$T_{\rho_0^\alpha} = 0.1560$, 
$T_{\tau_{\text{Peierls}}} = 0.06172$, 
$T_{\nu_0} = -0.03663$. 
Since the ranking does not change when the quantity of interest changes from $\varepsilon_\text{Y}$ to $\sigma_\text{Y}$, we conclude that the importance of parameters is $p > \Delta H_0 > C_{\lambda} > q > \rho_0^\alpha > \tau_\text{Peierls} > \nu_0$.

\section{Discussion}
\label{sec:discussion}

In this paper, we conducted several UQ studies for constitutive models in CPFEM with a single microstructure RVE for each case study. 
Three case studies are performed with different crystal structures, namely fcc, hcp, and bcc, for Cu, Mg, and W, respectively. 
\textcolor{black}{In this paper, three materials systems with different crystal structures (fcc, bcc, and hcp) are studied. Depending on the crystal structure, there may be different slipping and twinning systems in terms of slipping and twinning directions in plastic deformation, leading to interesting materials behaviors and mechanisms.} 
The quantities of interest are selected as the initial yield strain $\varepsilon_\text{Y}$ and the initial yield stress $\sigma_\text{Y}$. 
For fcc Cu, parameter $\tau_0$ is the most influential parameter for the initial yield behavior. 
For hcp Mg, all $\tau_{0,\text{basal}}$, $\tau_{0,\text{C}2}$, and $n_\text{tw}$ parameters are influential for the initial yield behavior. 
For bcc W, $p$ parameter in the short-range activation energy model is the most influential parameter for the initial yield behavior. 
% \chhighlight{I THINK IT WOULD BE GREAT TO MENTION WHICH PARAMETERS WERE MOST INFLUENTIAL IN EACH CASE.}

UQ studies, such as those described in this manuscript, play an important role in constitutive model calibration for unknown material system in the future. 
Since there are only a limited number of physical constitutive models, it is important to conduct a UQ study to observe the range of quantities of interest, and to numerically rank the influence of constitutive model parameters. 
Based on the stress-strain compilation curve conducted for various constitutive model parameters, the material behaviors can be rigorously quantified. 
The obtained UQ results provide a foundational step for further constitutive model calibration for future works, mostly conducted via digital image correlation techniques~\cite{turner2015digital,reu2018dic,reu2021dic}. 

Compared to polynomial approximation with full tensor grid, sparse grid approaches have a significant computational advantages, where this advantage grows with increasing dimensionality \textcolor{black}{thanks to a slower growth rate}~\cite{nobile2008sparse}. 
In the context of constitutive models, the computational reduction is mostly profound in the case of hcp system (as opposed to bcc and fcc), such as Mg and Ti, and in the case of dislocation-density-based constitutive model (as opposed to phenomenological model),  where many parameters require careful calibration to obtain a sufficient agreement with experiments. 
For simple system with a relatively simple phenomenological constitutive model, the computational reduction is less severe. 
\textcolor{black}{
It is noteworthy that the level of the Smolyak sparse grid in this study has a little effect on the resulting probability density function of QoIs. 
This implies that the underlying function is perhaps mostly low-order. 
This observation can also be confirmed by the Sobol' indices, where the first-order Sobol' indices are much more dominant, compared to higher-order Sobol' indices. 
}

% {\color{red} TMW: I don't see how the following paragraph is relevant to this paper.} % AT: it's very closely related topic -- optimization under uncertainty where uncertainty is quantified in this paper. Exactly how to do that remains an open question, but probably not by PCE or SC. So I made a comment of forward UQ may be helpful for optimization under uncertainty works in the future. 

To construct the response surface model, stochastic collocation provides a significant advantage for reducing the curse of dimensionality. 
However, when it comes to accuracy, Gaussian process regression, which is also the underlying surrogate model for Bayesian optimization, is arguably one of the best approaches in shallow machine learning. 
The direction of coupling Bayesian optimization, e.g.~\cite{tran2019pbo,tran2020smfbo2cogp,tran2021srmo,tran2019aphBO2GP3B,tran2022aphbo}, for robust constitutive model calibration remains open for future research. 

\textcolor{black}{
The scope of this manuscript is to quantify the microstructure-sensitive uncertainty. Obviously, it can be expanded to account for the entire stress-strain curve. However, due to the number of parameters involved in each constitutive model, there are hundreds to thousands of runs needed for a single microstructure RVE. Such computationally expensive numerical experiments require careful planning and execution, and therefore, remain a potential topic for future studies.
}
It is important to point out that by restricting to one RVE per case study, this work does not address microstructure-sensitive uncertainty that either is related or induced by the underlying stochastic nature  of microstructures. 
The direction of investigating a microstructure ensemble with many RVEs remain open for future work.

\section{Conclusion}
\label{sec:conclusion}

In this paper, we applied SC to quantify uncertainty associated with the initial yield behavior, mainly the estimated yield strain $\varepsilon_\text{Y}$ and the estimated yield stress $\sigma_\text{Y}$ for fcc Cu, hcp Mg, and bcc W. 
A variety of constitutive models are used, resulting with different parameterization and dimensionalities for the constitutive models considered. 
To mitigate the curse of dimensionality, Smolyak sparse grid is employed for high-dimensional integration to evaluate the PCE coefficients. 
Variance-based global sensitivity analysis is used to study the sensitivity analysis of the constitutive model parameters. 

\textcolor{black}{In light of the computational results presented in previous sections, there are several influential parameters that may have a significant effect on the initial yield behavior. For the phenomenological constitutive model, the slip resistance $\tau_0$, the slip hardening parameter $h_0$, and the strain rate sensitivity parameter $n$ are the most influential parameters, ranking in descending order. For the dislocation-density-based constitutive model, the $p$-exponent in glide velocity is the most influential parameter, followed by the activation energy for dislocation glide $\Delta H_0$, the dislocation mean free path parameter $C_{\lambda}$, the $q$-exponent in glide velocity, and the initial dislocation density $\rho_0^{\alpha}$. We conclude that in both constitutive models considered in this study, i.e. phenomenological (with and without twinning) and dislocation-density-based constitutive models, regarding the initial yield behavior, some parameters may have a profound effect on the QoI, while some others may not have a significant effect. The observation could potentially pave way for dimensionality reduction in constitutive model calibration in the future.
}

% {\color{red} TMW: This is more like a summary than a conclusion.} % AT: I agree. The conclusion has been expanded to complete the paper.

\section*{Conflict of Interest Statement}
%All financial, commercial or other relationships that might be perceived by the academic community as representing a potential conflict of interest must be disclosed. If no such relationship exists, authors will be asked to confirm the following statement: 
The authors declare that the research was conducted in the absence of any commercial or financial relationships that could be construed as a potential conflict of interest.

\section*{Author Contributions}

A.T. conceptualizes, performs simulations, analyzes data, and drafts the manuscript. B.T. and T.W. consults in uncertainty quantification aspect. H.L. consults in crystal plasticity finite element aspect. 

% The Author Contributions section is mandatory for all articles, including articles by sole authors. If an appropriate statement is not provided on submission, a standard one will be inserted during the production process. The Author Contributions statement must describe the contributions of individual authors referred to by their initials and, in doing so, all authors agree to be accountable for the content of the work. Please see  \href{http://home.frontiersin.org/about/author-guidelines#AuthorandContributors}{here} for full authorship criteria.

% \section*{Funding}
% Details of all funding sources should be provided, including grant numbers if applicable. Please ensure to add all necessary funding information, as after publication this is no longer possible.

\section*{Acknowledgments}
% This is a short text to acknowledge the contributions of specific colleagues, institutions, or agencies that aided the efforts of the authors.

The views expressed in the article do not necessarily represent the views of the U.S. Department of Energy or the United States Government. Sandia National Laboratories is a multimission laboratory managed and operated by National Technology and Engineering Solutions of Sandia, LLC., a wholly owned subsidiary of Honeywell International, Inc., for the U.S. Department of Energy's National Nuclear Security Administration under contract DE-NA-0003525.

% \section*{Supplemental Data}
% \href{http://home.frontiersin.org/about/author-guidelines#SupplementaryMaterial}{Supplementary Material} should be uploaded separately on submission, if there are Supplementary Figures, please include the caption in the same file as the figure. LaTeX Supplementary Material templates can be found in the Frontiers LaTeX folder.

\section*{Data Availability Statement}

% The datasets for this study is not available due to security concerns from Sandia National Laboratories.
The datasets for this study are available upon reasonable request.
% The datasets [GENERATED/ANALYZED] for this study can be found in the [NAME OF REPOSITORY] [LINK].
% Please see the availability of data guidelines for more information, at https://www.frontiersin.org/about/author-guidelines#AvailabilityofData

\bibliographystyle{frontiersinSCNS_ENG_HUMS} % for Science, Engineering and Humanities and Social Sciences articles, for Humanities and Social Sciences articles please include page numbers in the in-text citations
% \bibliographystyle{frontiersinHLTH&FPHY} % for Health, Physics and Mathematics articles
% \bibliography{test}
\bibliography{lib}

\end{document}